%% file: rsitemplate_AHS_aug_13.tex
\documentclass{revtex4-1} 
\usepackage{hyperref}
\usepackage[bottom]{footmisc}
\usepackage{url}
\usepackage{csquotes}
\usepackage{graphicx}
\usepackage{dcolumn}
\usepackage{bm}
\usepackage[english]{babel}
\usepackage{setspace}

\draft 

\begin{document}

\title{Enhancing the Performance of the T-Peel Test for Thin and Flexible Adhered Laminates} 

\author{Nikhil Padhye}
 \email{npdhye@gmail.com}
\author{David M. Parks}%
 \email{dmparks@mit.edu}
\affiliation{Department of Mechanical Engineering\\Massachusetts Institute of Technology}
\author{Alexander H. Slocum}
 \email{slocum@mit.edu}
\affiliation{Department of Mechanical Engineering\\Massachusetts Institute of Technology}
\author{Bernhardt L. Trout}
 \email{trout@mit.edu}
 \affiliation{ Department of Chemical Engineering\\Massachusetts Institute of Technology}

\begin{abstract}

Symmetrically bonded thin and flexible T-peel specimens, when tested on vertical travel machines, can be subject to
significant gravitational loading; with the associated asymmetry and mixed-mode failure 
during peeling.  This can cause erroneously high experimental peel forces to be recorded which leads to uncertainty in estimating
interfacial fracture toughness and failure mode. To overcome these issues, a mechanical test fixture has been designed for use with vertical test machines, that supports the unpeeled portion of the 
test specimen and suppresses parasitic loads due to gravity from affecting the peel test. 
The mechanism, driven by the test machine
cross-head, moves at one-half of the velocity of the cross-head 
such that the unpeeled portion always lies in the plane of the instantaneous center of motion. 
Several specimens such as 
bonded polymeric films, laminates, and commercial tapes were tested with and without the fixture, and the 
importance of the proposed T-peel procedure has been
demonstrated \footnote{The supplementary videos can be obtained from \url{http://web.mit.edu/npdhye/www/supplementary-videos.html} or by emailing npdhye@gmail.com.
For details on adapting or using the T-peel fixture for testing applications contact npdhye@gmail.com }. \\

\noindent \textit{Keywords: Peel Test, Mixed-Mode Fracture, Machine Design.}\\
 
\end{abstract}

\maketitle 

\noindent \textbf{Notation}

\noindent P is the force applied by the upper and lower grip in the symmetrical T-peel test schematic [N]\\
P' is the vertical force applied by the upper grip in the asymmetrical T-peel test schematic [N]\\
P'' is the vertical force applied by the lower grip in the asymmetrical T-peel test schematic [N]\\
P$_t$ is the vertical force per unit width applied by the upper grip on the upper peel arm[N/m]\\
P$_o$ is the reaction force per unit width at the base on the upper peel arm [N/m]\\
M$_t$ is the moment per unit width applied  by the upper grip on the upper peel arm [N]\\
M$_o$ is the reaction moment per unit width at the base of the upper peel arm [N]\\
M$_f$ is the moment applied by the upper grip on the upper peel arm in the symmetric T-peel test [N$\cdot$m] \\
M$_b$ is the moment exerted on the upper peel arm near crack tip in the symmetrical T-peel test schematic [N$\cdot$m]		\\
$\dot{x}$ is the rate of advance of crack tip	[m/sec] \\
$\sigma_x$ is the normal stress due to bending \\
$\theta$ is the angle made by the tangent at any point along the peel arm with respect to horizontal [rad] \\
$\sigma_{\mathrm{yield}}$ is the yield strength of the material in tension [Pa]\\
$\nu$ is the Poisson's ratio \\
E is the elastic modulus [Pa] \\
E'=$\frac{E}{1-\nu^2}$ is the plane strain elastic modulus [Pa]\\
t is the thickness of one peel arm [m]\\
w (or b) denotes the width of the peel specimen into the plane [m]\\
s is the coordinate along the peel arm [m] \\
I$=\frac{wt^3}{12}$ is the moment of inertia of beam out of plane [m$^4$] \\
$\rho=\frac{d s}{d\theta}$ is the radius of curvature at any point on the beam [m]\\
$\kappa=\frac{1}{\rho}$ is the curvature at any point on the beam [m$^{-1}$]\\
U is the bending elastic energy [J]\\
G$_c$ is the total critical energy release rate [J$\cdot$m$^{-2}$]	\\
G$_{Ic}$ is the critical energy release rate in mode I failure [J$\cdot$m$^{-2}$]	\\
G$_{IIc}$ is the critical energy release rate in mode II failure [J$\cdot$m$^{-2}$]	\\
G$_{IIIc}$ is the critical energy release rate in mode III failure [J$\cdot$m$^{-2}$]	\\
k$=\frac{P}{E'I}$ is a user defined constant [m$^{-2}$]\\
$P_1^T$	 is the vertical force on the upper grip in the asymmetric T-peel test [N]				\\
$P_1^B$	 is the vertical force on the lower grip in the asymmetric T-peel test	[N]				\\
$P_2^T$	 is the horizontal force on the upper grip in the asymmetric T-peel test	[N]			\\
$P_2^B$	 is the horizontal force on the lower grip in the asymmetric T-peel test	[N]			\\
$M^T$ is the moment at the upper grip in the asymmetric T-peel test			[N$\cdot$ m]	\\
$M^B$	is the moment at the lower grip in the asymmetric T-peel test			[N$\cdot$ m]	\\
$\psi$ is the phase angle during mixed-mode cracking [rad]\\
$K_I$ is mode I stress intensity factor	[Pa$\sqrt{m}$]				\\
$K_{II}$ is mode II stress intensity factor	[Pa$\sqrt{m}$]				\\
$W$	 is the weight of the peel specimen 	[N]			\\
$V_1$ is the internal shear force  parallel to the surface of upper peel arm			[N]			\\
$V_2$ is the internal shear force  parallel to the surface of lower peel arm			[N]			\\
$F_1$ is the internal normal force  on the surface of upper peel arm			[N]		\\
$F_2$ is the internal normal force on the surface of lower peel arm		[N]		\\
$M_{1}$ is the internal moment acting on upper peel arm 				[N$\cdot$ m]\\
$M_{2}$ is the internal moment acting on lower peel arm 				[N$\cdot$ m]\\
$M_g$ is the moment due to weight on the tail of the specimen		[N$\cdot$ m]\\
$M'$ is the internal moment on the tail of the specimen [N$\cdot$ m]\\
$M_{1g}$ is the internal moment in reaction to gravity [N$\cdot$ m]\\
$M_{2g}$ is the internal moment in reaction to gravity [N$\cdot$ m]\\
X, X', and X'' are the distances of different sections of peel specimen from line of 
action of vertical forces		[m]\\
X$_{max}$ is the maximum span of the peel arm in x-direction [m]						\\	
Y$_{max}$ is the maximum span of the peel arm in x-direction [m]						\\
Y$_1$ and Y$_2$ are the distances of the upper and lower grips from the shown section of the peel specimen		[m]\\	
$L$ is the length of the peel arm	[m] \\
$\Gamma$ effective fracture surface energy 		[N/m]			\\
$\delta$ is the irreversible work per unit area of crack advance 		[N/m]			\\
$\rho$ density of the polymer			[kg/m$^3$]		\\
$\sigma_{22}$ normal stress in the vicinity of the crack tip								[Pa]\\
$\sigma_{12}$ shear stress in the vicinity of the crack tip							[Pa]\\

\section{INTRODUCTION}

The peel test is a common simple mechanical test for measuring adhesion strength in various applications, especially bonded thin and flexible laminates, 
and it can be performed in several different ways. The procedure usually consists of a laminate bonded to another laminate or to a thick substrate, and the test is conducted by pulling 
the laminate off the laminate or substrate at some angle while recording the peeling force in the steady-state during debonding. 
The usual goal is to relate the experimentally-obtained peel force to the 
intrinsic fracture toughness of the interface, where the interface toughness represents 
the work required per unit area to advance a crack at the interface and has the units J/m$^2$ or N/m. 
Only in restricted scenarios can the peel force give a direct estimate of interface toughness. 
In general, the peel force is affected by the sample geometry, constitutive properties 
of the laminates, interfacial properties, and the test apparatus.  

Depending on the application, several existing ASTM \cite{Astmd903,Astmd3807,AstmdD6252,ASTMD1876,ASTMD3330,
ASTMD6862,ASTMF88} or ISO standards \cite{ISO11339,ISO8510,ISO11339,
ISO14676,ISO29862} are commonly employed for measuring
adhesion through peeling. These standards can be easily practiced on
various commercially available fixtures, such as 90$^\circ$, 180$^\circ$, climbing drum, floating roller, adjustable angle, and 
German rotating wheel peel fixtures \cite{Admet-web,InstronWebsite4,MecmesinWebsite}. 

Amongst the earliest attempts, two approaches for analyzing peeling were adopted: (i) energy based criterion, and (ii)
stress based criterion. 

The energy 
approach can be traced to work by Rivlin 
where strip detachment under the effect of weight was investigated.
The stress-based criteria assumed that the failure occurred when the tensile stress in the 
adhesive interlayer exceeded the tensile strength due to peeling. 
The first theoretical analysis of peeling was presented by Spies \cite{spies1953peeling}.
Here, the non-bonded part of the peeling strip was treated as an elastica,
and the bonded part of the peeling strip was treated as an elastic beam on an elastic (Winkler) 
foundation of the adhesive layer. The failure criterion was based on the tensile stress exceeding
the tensile strength of the adhesive interlayer. This approach implicitly assumed that rupture did not occur at any of 
the interfaces but rather in the bulk of the adhesive; the adhesive was assumed to behave as a Hookean solid up to the point of rupture, 
and stress concentrations at the strip edges were disregarded. Spies \cite{spies1953peeling} recognized the 
role of plastic deformation during peeling strip and used an average 
elastic constant to account for the ductile behavior of the strips. 

Since then, several investigations similarly treated the unbonded part of the flexible laminate 
as an elastica, whereas the bonded part as an elastic beam on an elastic foundation. Bikerman \cite{bikerman1957theory} 
considered 90$^\circ$ elastic peeling of the adherend with explicit solutions 
for Newtonian interlayer flow as well as cohesive failure of a Hookean interlayer.
Jouwersna \cite{jouwersma1960theory} presented a refinement over Bikerman's model. 
Kaelble \cite{kaelble1959theory} considered elastic peeling and emphasized the 
roles of cleavage and shearing mechanisms of failures during peeling and their effect on the peel force,
and later reported rate effects in peeling \cite{kaelble1960theory}.
Yurenka \cite{yurenka1962peel} considered the peeling of the adhesively bonded metal, and
analyzed several peeling configurations such as an edge bending load peel test,
T-peel test, four-inch diameter drum peel test, four foot diameter drum peel test, 
and climbing drum peel test. Nicholson \cite{nicholson1977peel} considered the effects of large bending deformations of the beam 
during peeling.

The condition for the fracture of an adhesive joint during peeling using an energy balance (or energy theory of fracture) was
presented by Kendall \cite{kendall1971adhesion}. Subsequently,
Kendall studied the rate effects during viscoelastic peeling \cite{kendall1973peel}
and the dependence of the peel force on adhesive surface energy and elastic deformation of the adherend \cite{kendall1973shrinkage,kendall1975thin}.
Chang et al. \cite{chang1960peeling} employed a linear viscoelastic model to describe the constitutive behavior of the adherend and 
presented approximate solutions for several peeling configurations.
Chen et al. \cite{chen1972mechanics} were the first to consider the elastic-plastic behavior in peeling 
but used the tensile strength criterion for modeling the adhesive failure. 
Chang et al. \cite{chang1972effects} considered an elastic-perfectly plastic analysis for adhesive failure, and
accounted for the energy dissipation during plastic bending.  

Gent et al. \cite{gent1975peel} highlighted the difference between
the energy-based and stress-based criteria for failure in peeling, and they indicated that the plastic deformation 
of the flexible members during peeling could account for an enhanced peel force. In another
study  \cite{gent1977peel}, effects of plastic deformation in 90$^\circ$ and 180$^\circ$
were considered, and an approximate analysis of elastoplastic peeling was presented.
A numerical solution of elastoplastic peel problem 
has been given by Crocombe and Adams \cite{crocombe1981peel}, and 
stress fields ahead of the crack tip were later studied \cite{crocombe1982elasto}. The role of 
residual strain energy in elastoplastic peeling was emphasized by Atkins et al. \cite{atkins1986residual}. 
Kim et al. \cite{kim1988elastoplastic,aravas1989mechanics} comprehensively studied the elastoplastic peeling from a rigid substrate, 
and estimated the energy dissipation due to inelastic bending in the peel arm. 
To estimate the true interfacial fracture energy, a correction scheme of subtracting the structural dissipation from 
total external work was proposed. In related studies \cite{kim1988elasto,kim1989mechanical}, similar elastoplastic peeling 
of the adherends was considered, energy losses due to viscoelasticity were analyzed, and experimental results were reported.
Kinloch et al. \cite{kinloch1994peeling} employed beam theory models to investigate the elastoplastic deformation in peeling,
and they analyzed the role of finite beam rotation at the base of peeling arm.

Wei and Hutchinson \cite{wei1998interface} were the first to report cohesive zone stress effects on
the plastic deformation of the adherends near the peeling front and its effect on
the elevated levels of peel forces. It was shown that, if the maximum cohesive stress along the interface was
greater than about three times the yield stress of the laminate, the steady-state peel force considerably increased. 
Larger cohesive stresses lead to a dominant tri-axial stress state near the peeling front, and
a simple beam model is no longer appropriate to describe the deformation of the adherend. 
Yang et al.  \cite{yang1999numerical,yang2000analysis} studied the T-peel
while accounting for excessive plastic deformation in the adhesively bonded laminates,
and they used embedded process zone models for fracture. Thouless et al. \cite{thouless1992elastic} studied 
the mixed-mode effects during the peeling of an elastic adherend from an elastic substrate. 
In addition, De Lorenzis et al. \cite{de2008modeling} employed finite element analysis for mixed-mode debonding of laminates 
from rigid substrate in the peel test.
Yang et al. \cite{yang2001mixed} analyzed the mixed-mode fracture 
of plastically deforming adhesive joints. Williams et al. \cite{williams2002analytical} 
employed cohesive zone models and presented solutions for the peeling of a cantilever beam specimen bonded to an elastic foundation.
Su et al. \cite{su2004elastic} developed an elastic-plastic interface constitutive model
and successfully applied it to various geometries such as T-peel test, four-point bend experiments on bonded 
bi-layer edge notch specimens, and lap shear experiments. 
Zhang et al. \cite{zhang2009generalized} also developed numerical models for adhesives. 
Hadavinia et al. \cite{hadavinia2006numerical} performed a 
numerical analysis of elastic-plastic peel tests (90$^\circ$ and T-peel) and tapered double cantilever 
beam specimens. 

\begin{figure}[!htb]
\centering
\includegraphics[scale=0.6]{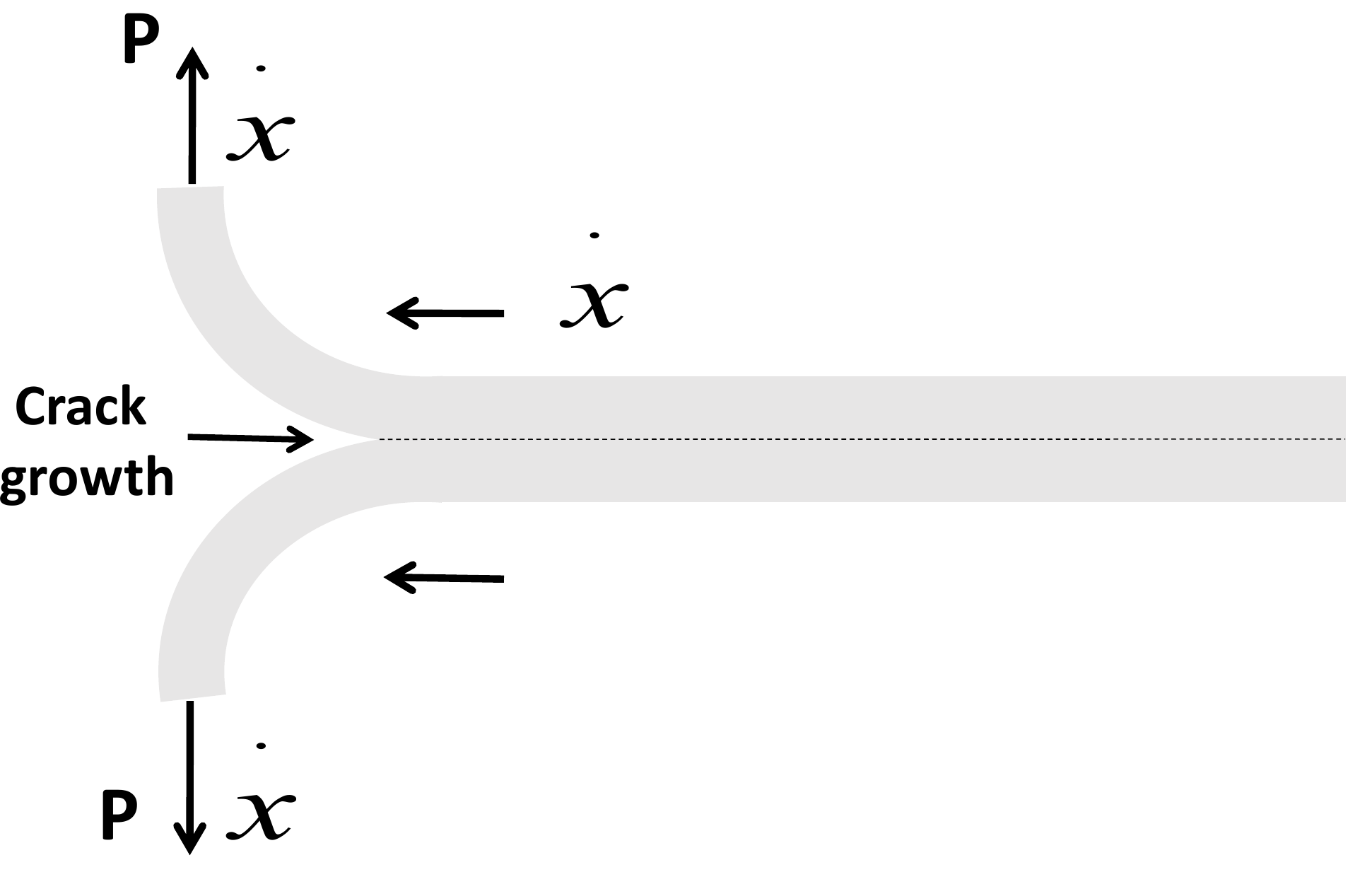}
\caption{Ideal T-peel test with symmetric loading and material properties across the interface. }
\label{fig:peel-test}
\end{figure}

This study deals with a particular type of peel test, known as a \textit{T-peel test}
(\textit{T} indicating that the specimen forms a T-shape) between symmetrical interfaces.
The T-peel test has been practiced for at least the past six decades \cite{deBruyne1951}.  
For discussions, we shall refer to an \textit{ideal} T-peel test which assumes the following:

(i) The two symmetrical halves are joined together with an infinitesimal (or nearly zero thickness) intermediate layer.

(ii) Crack propagation occurs on a predefined path between the two symmetrical halves through the intermediate layer.

(iii) Under symmetrical loading, the two halves separate purely by opening with G$_{Ic}$ as mode I fracture 
energy per unit area of crack propagation.

(iv) The peeling force during the test is not affected by gravity. 

Figure ~\ref{fig:peel-test} shows an idealized T-peel scenario with symmetric loading and material properties across the interface. 
We shall frequently drop the letter T, but the reference to the T-peel test is implied.  
During the steady-state crack propagation, if `P' is the debonding load (N), `w' indicating the specimen width (m) 
into the plane, and $\dot{x}$ is the velocity (m/s), then the rate of total external work is given by:

\begin{equation}
\label{eq:work-energy-balance}
W_{\mathrm{external}} = 2P\times \dot{x}
\end{equation}

From the work energy balance, under the assumptions of quasi-static (no kinetic energy) and steady-state conditions (elastic energy stored in the peel arm is constant), 
the rate of total external work ($W_{\mathrm{external}}$) during the crack propagation equals to the sum of rate of work due to interface debonding and any dissipation 
associated with the specimen deformation (say, plastic deformation due to bending of arms, discussed in detail
in Section ~\ref{sec:slender-beam-theory}). 
If $\gamma$ and $\delta$ denote the debonding and dissipative energies, respectively, per unit area of crack advance, then

\begin{equation}
\label{eq:work-energy-balance-2}
W_{\mathrm{external}} = 2P\times \dot{x} =   (\gamma+\delta) \times w \times \dot{x}
\end{equation}

Therefore, 
\begin{equation}
\gamma + \delta = \frac{2P}{w}
\end{equation}

Note, $\delta$ refers to the dissipative work in the specimen other than within the interfacial region, and the effect of crack tip inelasticity is included in $\gamma$,
i.e., $\gamma$ represents the sum of the reversible and associated irreversible work necessary for crack
advance at the interface \cite{irwin1958fracture,orowan-fantastic-1946}. 
A clear distinction between $\delta$ and $\gamma$ may not always be straightforward
in the scenarios when the regions of structural plasticity and crack tip plasticity overlap.  

$\delta$ as well as $\gamma$ can depend on numerous factors including test conditions and material constitutive properties,
but they attain constant values during steady-state peeling. 
If $\delta$ is not equal to zero
then we must subtract the work corresponding to $\delta$ from 
$W_{\mathrm{external}}$, 
to arrive at the \textit{correct} interface toughness  ($\gamma$). 
If  $\delta=0$ then the peel force gives a direct measurement of the 
interfacial fracture energy. In particular, if $\delta$ is zero, and
symmetrical loading and material properties prevail 
on both sides of the interface such that the failure of the 
interface occurs purely by opening mode, then $\gamma (=\frac{2P}{w}$) represents the mode I fracture  
toughness (G$_{Ic}$) of the interface. 
However, if the peel force exceeds the elastic limits, then failure is accompanied with plastic deformation in the structure, which 
is intrinsically not connected to the interfacial fracture and therefore $\delta$ is not 
equal to zero (equation ~\ref{eq:work-energy-balance-2}). In such cases, the peel force is not a direct measure of 
adhesive fracture energy. 

\begin{figure}[!htb]
\centering
\includegraphics[scale=0.6]{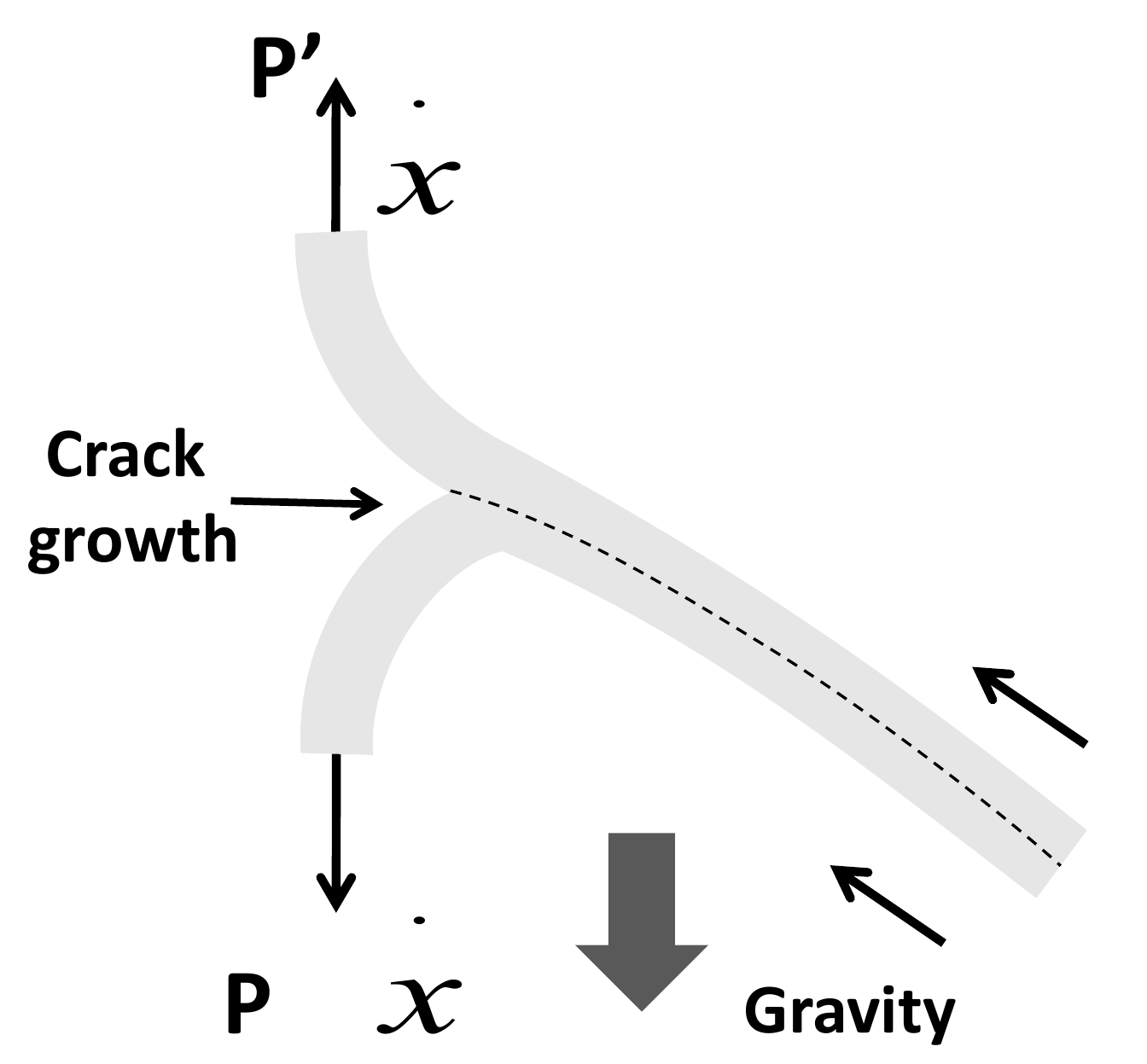}
\caption{Peel test in vertical test machines and the effect of gravity.
If the peel specimen has low bending stiffness then it can undergo large rotations and deformations under its own weight. }
\label{fig:peel-test-gravity-hangover}
\end{figure}

The T-peel test is quite straightforward to perform: the unbonded parts of the two flexible laminates are clamped in the grips of a mechanical tester
and separated. If the test is performed on a vertical mechanical tester,
the bottom grip is usually held fixed, whereas the top grip moves upwards.
In this vertical configuration, gravity can 
have following effects: 
(i) a contribution to the measured peel force due to specimen's weight, 
and (ii) an asymmetric T-peel configuration (particularly when a specimen
is thin and flexible). This is shown schematically in Figure ~\ref{fig:peel-test-gravity-hangover}. 
In this situation, the contribution to peel force to specimen weight
and, the bending of the freely suspended end and the
degree of anti-symmetry introduced, depend on the geometry (i.e. length, width, and thickness) 
and the material properties (i.e. density, modulus, etc.) of the specimens. 
One can imagine critical scenarios where the bending action of the gravity 
may induce plastic deformation in the lower peel arm.
All these effects could lead to a significant deviation in the measured peel force compared to the ideal T-peel. 
The uncertainty in the asymmetry during such a test is uncontrolled,
since the degree of mode-mixity (or phase angle) is unknown (discussed in section ~\ref{sec:mechanics-of-peeling-and-fracture}), 
and therefore no straightforward correction is possible for the same. 

Recently, there have been some studies on the role of asymmetry in T-peel specimens \cite{hirai2009performance,
choi2001peel}.  Dolah et al. \cite{dolah2014effect} have proposed a horizontal-pull testing apparatus to perform the T-peel in a layout such that the 
tail of the specimen aligns with the gravity. However, the mechanics of T-peel in this vertical-tail layout has not been discussed.

\begin{figure}[!htb]
\centering
\includegraphics[scale=0.55]{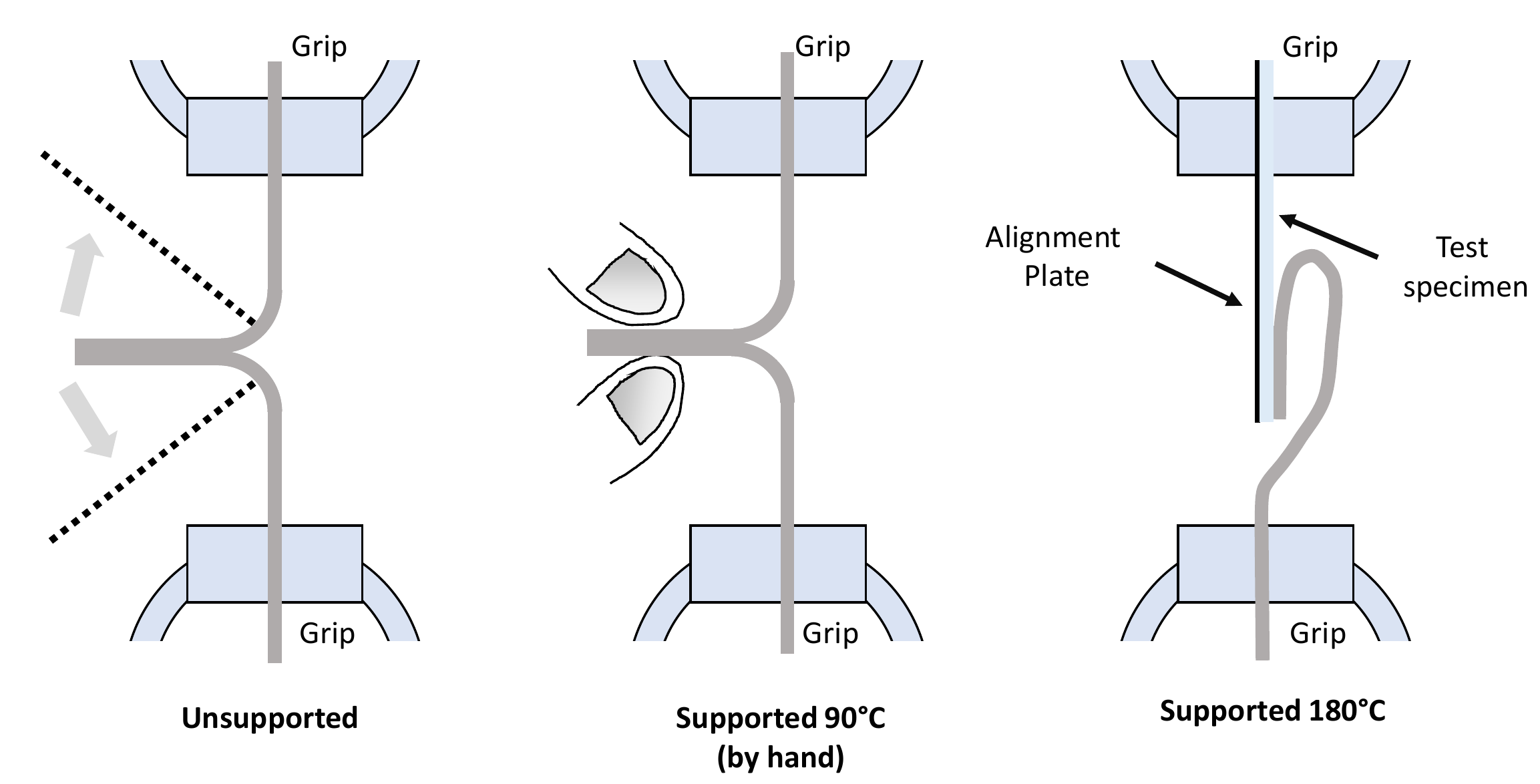}
\caption{T-peel test methods as described in the ASTM standard [18], and the recommended test adopted in medical packaging industry (MPT) is the unsupported method.}
\label{fig:peel-test-by-mpt}
\end{figure}

To the best of our knowledge, issues associated with T-peel testing of thin and flexible 
laminates on vertical test machines have not been resolved, and
no testing fixture has been reported to perform it with sufficient precision. 
Figure ~\ref{fig:peel-test-by-mpt} schematically shows 
a peel test procedure adopted by the medical packaging industry \cite{medical-packaging-industry},
where the ASTM F88/F88M-09 standard is also referred. In the ASTM standard F88/F88M-09 \cite{ASTMF88}, 
it is acknowledged that during a peel test a portion of the peel force may be 
due to bending component and not the ``strength'' alone, and that numerous fixtures and techniques have been devised 
to hold samples at various angles to the pull direction to control this bending force. However, the recommendation to compensate 
for the tail hangover during peel test is through hand support. It is quite evident that any such manual 
endeavor 
can itself introduce parasitic forces.

Even the commercial testing demonstrations \cite{InstronWebsite1,InstronWebsite2,InstronWebsite3} do not make any recommendation 
for avoiding the issues caused by the gravity. Instron$^{\textregistered}$
\cite{InstronWebsite3} proposes to use the ASTM D1876-08 \cite{ASTMD1876} standard for the peel test, 
and states that \enquote{this method caters to  the determination of the comparative peel or stripping characteristics of adhesives bonds when tested on standard-sized specimens 
under defined conditions of pretreatment, temperature, and testing speed. Typical results include \enquote{Average Load between 2 points}. 
This test can be performed on single column and dual column test frames.} The ASTMD 1876-08 peel test standard states that ``this test 
method is primarily intended for determining the relative peel resistance of adhesive bonds between flexible adherends by means of a T-type specimen 
using a tension testing machine. The bent, unbonded ends of the test specimen shall be clamped in the test grips of the tension test machine, 
and a load of a constant head speed shall be applied. An autographic recording of the load versus 
the head movement or load versus distance peel shall be made. The peel resistance over a specified length of the bond line after the initial peak shall 
be determined.''  No reference to account for specimen's weight, effects such as gravity-induced asymmetry and mixed-mode failure, plastic bending, or suggestion for using a fixture for more accurate testing are made.  Other examples of commercial testing procedures (where no reference to these effects has been made) can be found in \cite{video-youtube-1,video-youtube-2}. In \cite{Quality-Magazine}, common problems with tests in relation to adhesion are discussed, but no reference to issues related to effects on the T-peel test due to gravity is made. Similarly, in several other T-peel experimental studies on relatively thin laminates \cite{gardon1963peelII,gent1969adhesion,gent1994interfacial}, no reference to these effects is made. 

In addition to its simplicity, an ideal T-peel test can be employed to estimate the mode I fracture toughness ($G_{Ic}$) of thin and flexible laminates.  
However in other peel test procedures, such as 90$^\circ$ peel test (for e.g. ASTM D6252 standard \cite{AstmdD6252}), etc. 
when the symmetry across the bonded interface is broken; $G_{Ic}$ cannot be determined. 

To overcome limitations of current test methods, we have designed a mechanical fixture that suppresses the 
action of the gravity on T-peel tests performed on vertical test machines.
This study has been motivated by our recent work \cite{padhye-nature-bonding}, where authors
identified a compelling case such that the gravity showed a significant effect, and correct estimation of mode I fracture toughness 
$G_{Ic}$ was not possible. Next, we review the stress fields associated with the crack tip, 
mix-mode failures, energy release rates, and other concepts relevant to the peel test. 

 

\section{PEELING AND FRACTURE MECHANICS}
\label{sec:mechanics-of-peeling-and-fracture}
%

In case of two-dimensional geometries, when the material properties are symmetrical across the interface
and loading leads to pure opening displacements, then only normal stresses act across the plane ahead of the crack tip, and
fracture is referred to occur in mode I. The fracture toughness in such cases is normally denoted by $G_{Ic}$.

However, an asymmetry in the material properties or loading results in shear stresses ahead of the crack tip 
and a mixed-mode failure (comprising of both opening and shearing displacements). 
The fracture toughness ($G_{c}$) in a mixed-mode then depends upon the relative amounts of opening and shearing
deformations. A mixed-mode toughness is usually greater than pure mode I toughness ($G_{Ic}$).

As per linear elastic fracture mechanics,
a singular stress field is associated with the crack tip when
traction-free line crack exists. This can be expressed compactly by using
complex number notation in terms of
mode I ($K_I$) and mode II ($K_{II}$)
stress intensity factors as follows:\\

\begin{equation}
\label{eq:mixed-mode-stresses}
\sigma_{22}+i\sigma_{12} = \frac{ (K_I+iK_{II})}{\sqrt{2\pi r}},
\end{equation}
\\
In the above equation ~\ref{eq:mixed-mode-stresses}, $K_I$ and $K_{II}$  depend linearly on the applied loads and on the details of
the full geometry of the body. If ${E'}=E/(1-\nu^2)$ denotes the plane strain tensile modulus, then the total fracture toughness in terms of $K_I$ and $K_{II}$, as per Irwin's result, can be written as follows:\\

\begin{equation}
G_{c}= \frac{(K_I^2+K_{II}^2)}{E'}.
\end{equation}

The relative magnitudes of $K_{I}$ and $K_{II}$ are usually expressed in terms of
a phase angle $\psi$ as:\\

\begin{equation}
\psi=tan^{-1}(K_{II}/K_I).
\end{equation}
\\
The mixed-mode fracture toughness is usually a function of $\psi$
and it can be expressed as $G=G_c(\psi)$. Here, the subscript \textit{c} denots the
critical energy release rate during steady-state cracking. It is essential to emphasize 
that even in the presence of material symmetry, if an asymmetric loading occurs across the interface, then the interface toughness
will be a function of the phase angle $\psi$ (which will itself depend upon the overall loading configuration). 
Only when symmetric material properties and loading occur across the interface, can the mode I fracture toughness (G$_{Ic}$)
be estimated. For a detailed study on mixed-mode failures, the readers are referred to \cite{suo1990interface,hutchinson1990mixed}.

When T-peel tests are performed on vertical test machines with adherends exhibiting small bending stiffness, then 
gravity can introduce a major asymmetry. This leads to a mixed-mode failure.
In addition, the asymmetry can affect $\delta$ by
causing structural plasticity in the lower peel arm due to bending under its own weight. 
During the peel test, the overhanging tail end of the specimen shortens; therefore, 
the degree of asymmetry varies. Due to these variations, 
no straightforward correction is possible and the determination of the 
interface toughness is uncertain. We emphasize that 
even if we know the specimen's weight, there is no straight-forward
method to account for the uncertainties
associated with the asymmetric configuration.
See Appendix for details
on mechanics of symmetric and asymmetric T-peel test.

\section{DESIGN OF A T-PEEL TEST FIXTURE FOR VERTICAL CROSSHEAD TRAVEL MACHINES}
\label{sec:Desing-of-A-Peel-Fixture}

\begin{figure}[!htb]
\centering
\includegraphics[scale=0.5]{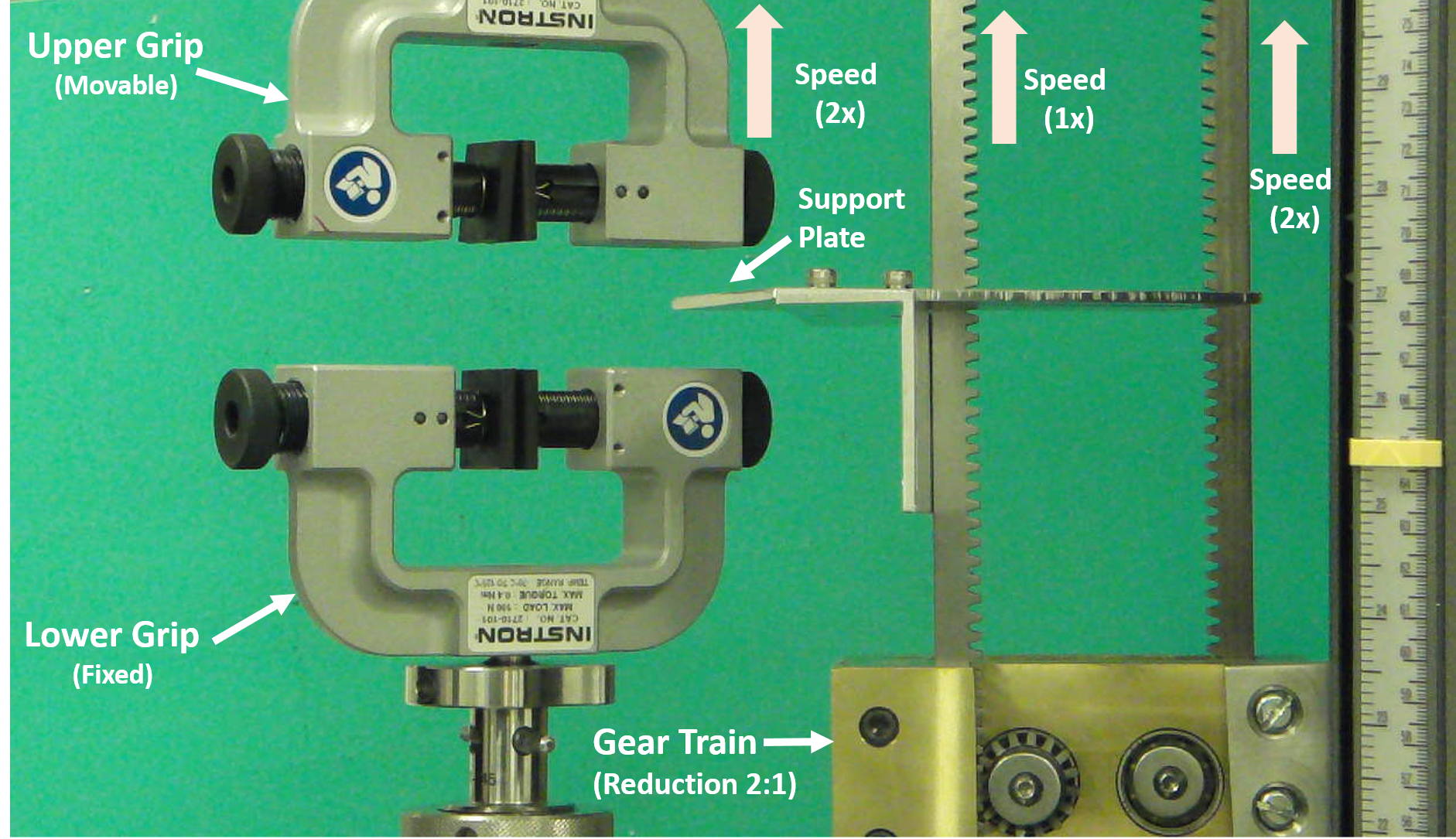}
\caption{Peel test fixture for a vertical machine. See supplementary video S1.}
\label{fig:peel-mechanism}
\end{figure}

During a T-peel test on a vertical machine, the two peel arms of the T-peel specimen are clamped in the upper and lower grips. 
In general, the lower grip is stationary, and the upper grip moves upwards. To support the specimen weight and 
maintain the symmetry of the T-peel specimen, 
a mechanism is needed that supports the freely hanging end of the specimen and moves upwards at a velocity 
half of that of the upper grip. Figure ~\ref{fig:peel-mechanism} shows such a mechanical fixture mounted on a 
vertical mechanical tester. 

\begin{figure}[!htb]
\centering
\includegraphics[scale=1.6]{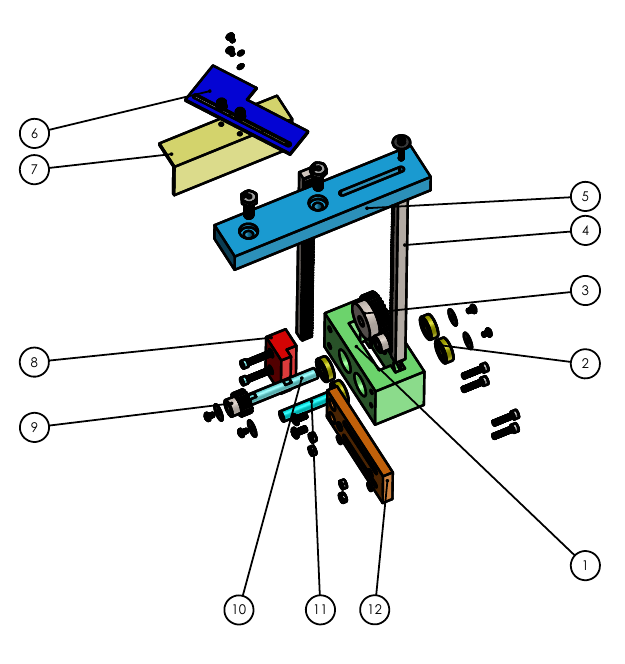}
\caption{Drawing of the peel mechanism assembly (elements listed in Table ~\ref{tab:bom}).}
\label{fig:NewGlobalCleanAssembly-reduced-size-cropped}
\end{figure}

The peel test fixture uses a 2:1 speed reduction to cause the platform that supports the specimen to move at half the speed of the moving cross-head.  
A driving rack that is attached to the cross-head of the test machine drives the first spur gear of diameter $D/2$.
The first spur gear of diameter $D/2$ is meshed with a second spur gear of diameter $D$. The second spur gear is mounted on the shaft which carries a third spur gear of diameter 
$D/2$. The third spur gear drives the driven rack; hence the driven rack moves at half the speed of the driving rack.
The driven rack is coupled to the plate that supports the overhanging tail of the peel specimen. 
As shown in the figure, the driving rack moves up at twice the speed of the driven rack. This causes the support plate attached to the driven rack to move 
at half the speed of the cross-head. We incorporated several holes/attachment points/or grooved slots on 
the driven rack such that the vertical and horizontal positions of the base plate could be adjusted before the start of the test.
The mechanical fixture and its solid model are
shown in Figure ~\ref{fig:NewGlobalCleanAssembly-reduced-size-cropped}. 
\begin{table}
 \begin{center}
 \caption{Bill of Materials for T-peel test Mechanism.}
 \label{tab:bom}
    \begin{tabular}{ | l | l |l|}
    \hline
\textbf{Item No.} 	&	\textbf{Part Description}	& 	\textbf{Qty.}\\ \hline \hline
  		1	 	& Gear Box					&		1		\\
  		2		& Bearing					&		4		\\
  		3	 	& Spur Gear (24 teeth)			&		1		\\
  		4	 	& Rack 						&		2		\\  	
  		5	 	& Top Support Hanger			&		1		\\
  		6	 	& Film Support Plate			&		1		\\
  		7	 	& Connecting Bracket			&		1		\\
  		8	 	& Rack Retainer				&		1		\\	
  		9	 	& Spur Gear (12 teeth)	 		&		2		\\
  		10	 	& Shaft (driven end)			&		1		\\  		
  		11		& Shaft (driving end)			&		1		\\  		
  		12	 	& Support Bracket				&		1		\\  		
    \hline
    \end{tabular}
\end{center}
\end{table}

From the experimental viewpoint, it is important to ensure that the crack front is sufficiently far from the nearest edge of 
the support plate. However, it should not be too far either, as the gravity-induced bending effects may appear on the
part of the specimen lying between the crack and the nearest edge of the support plate. In addition, one must ensure that 
the peel arms are not constrained by contact with the peel fixture.  If the frictional resistance arising from contact between the specimen tail and support plate were significant,
an air-bearing support can be introduced \cite{el2005frictionless}. However, in the current application where the tapes are very lightweight, 
the resulting effect of the frictional drag force would produce a fraction of a  percent error and thus we did not see a need for a frictionless platform support. 

\section{EXPERIMENTS, RESULTS AND DISCUSSIONS}
\label{sec:results-and-discussions}
\begin{figure}[!htb]
        \centering
                \includegraphics[scale=0.3]{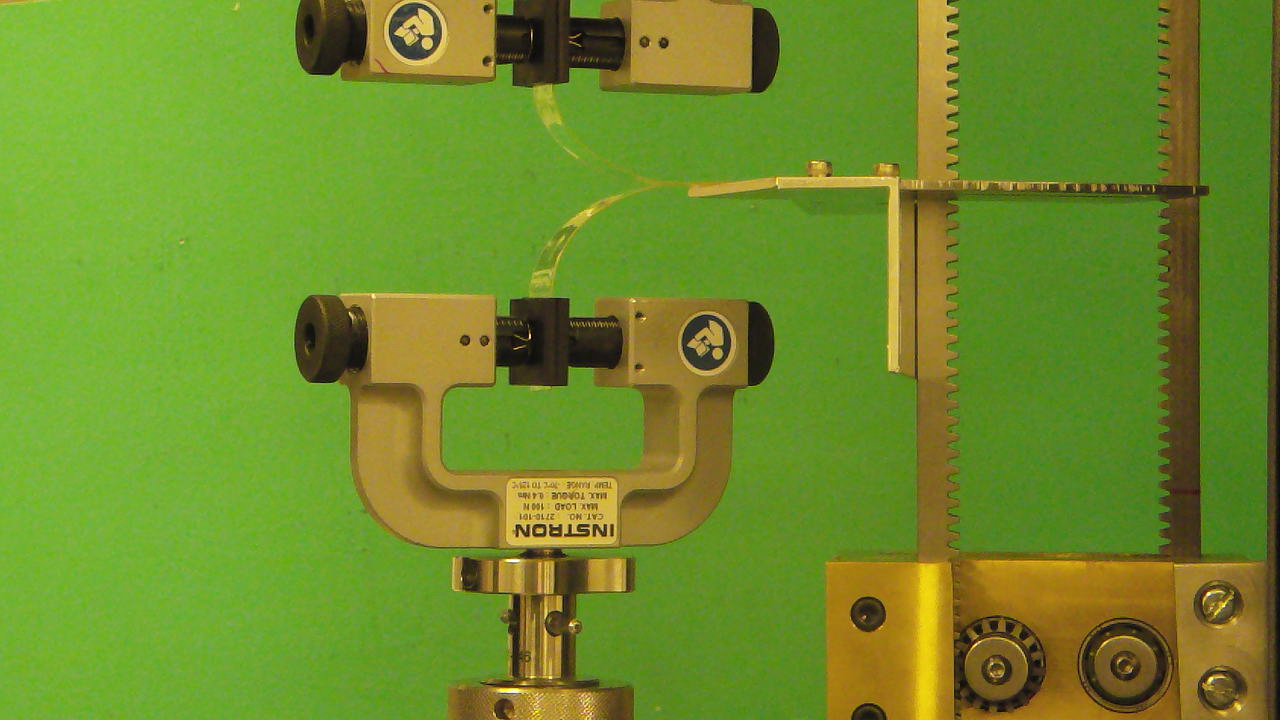}
                \caption{T-peel test on a vertical machine with the designed mechanical fixture. See supplementary video S2.}
                \label{fig:peel-test-with-mechanism}
\end{figure}
    \begin{figure}[!htb]
        \centering
                \includegraphics[scale=0.3]{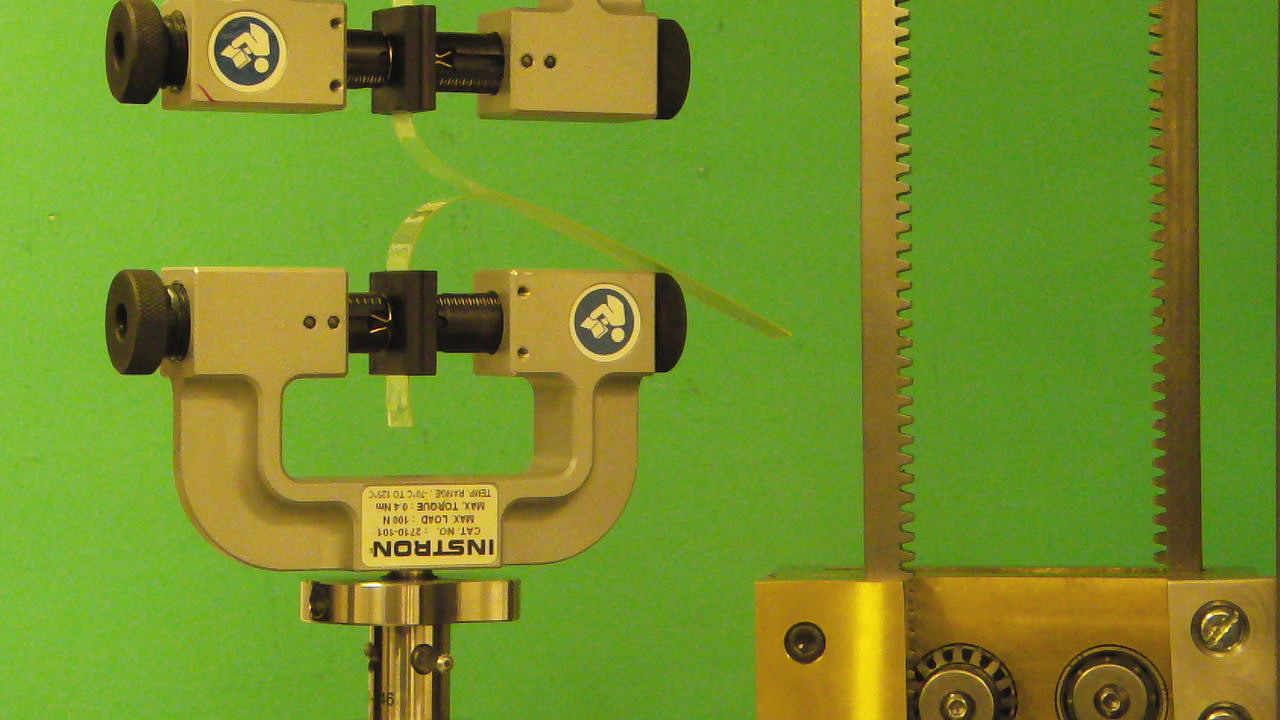}
                \caption{T-peel test on a vertical machine without the mechanical fixture. See supplementary video S3.}
                \label{fig:Without-mechanism-peel-test}
\end{figure}
\begin{figure}[!htb]
\centering
\includegraphics[scale=0.5]{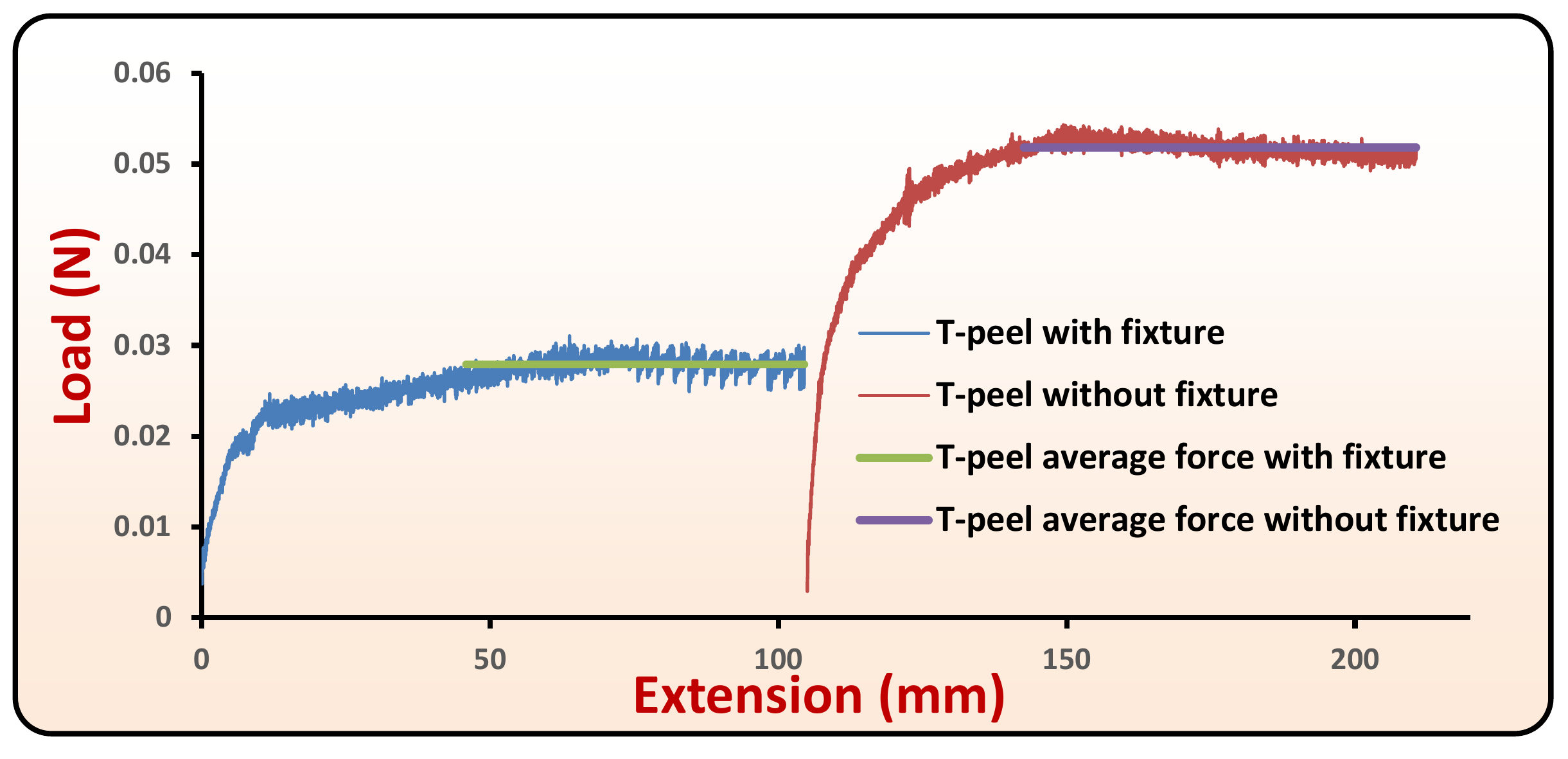}
\caption{T-peel test on a vertical machine with and without using the mechanical fixture for roll-bonded laminates sample `A'.}
\label{fig:t-peel-test-comparison}
\end{figure}
\begin{figure}[!htb]
\centering
\includegraphics[scale=0.5]{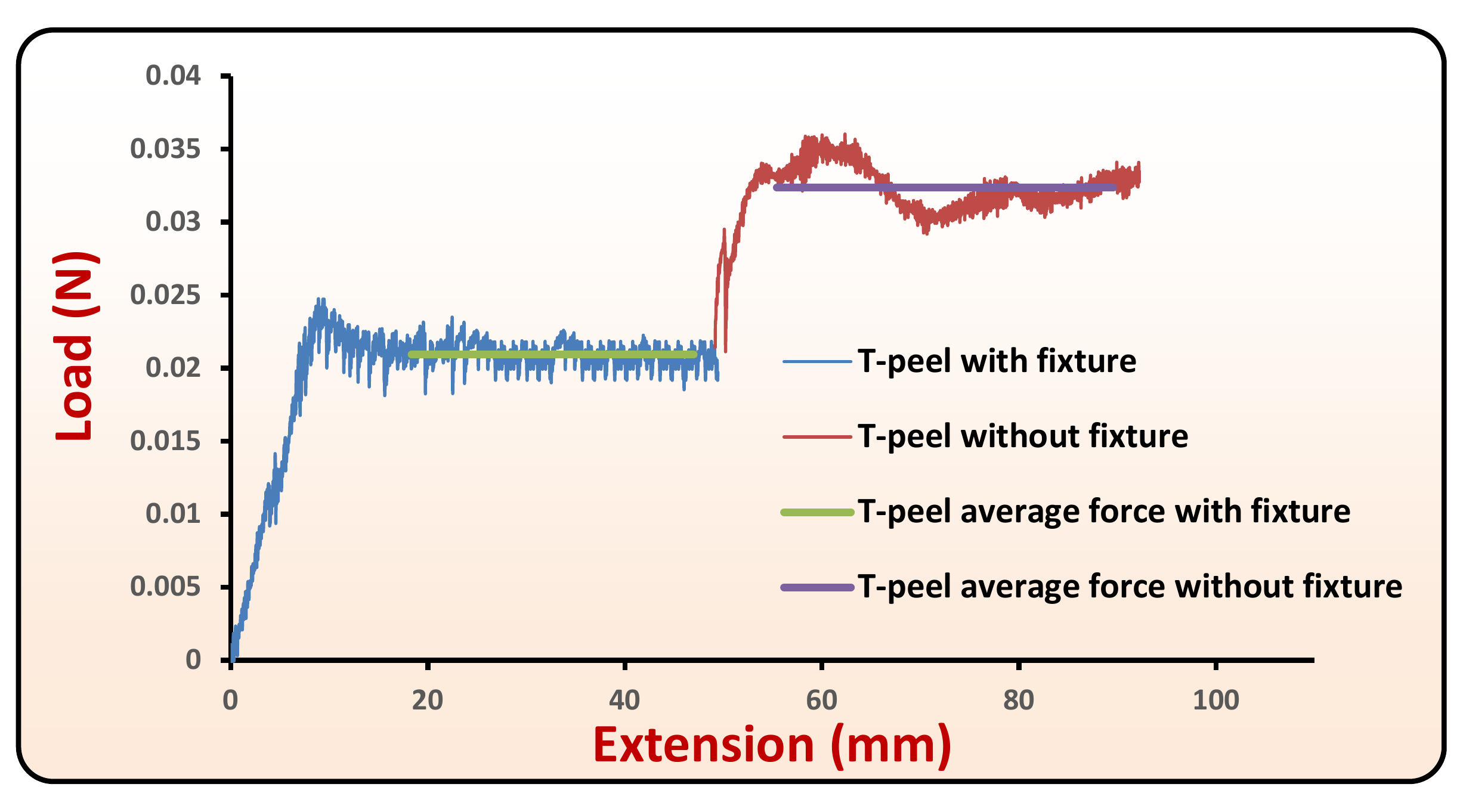}
\caption{T-peel test on a vertical machine with and without using the mechanical fixture for roll-bonded laminates sample `B'.}
\label{fig:S1-30th-aug-2015-cropped}
\end{figure}
\begin{figure}[!htb]
\centering
\includegraphics[scale=0.5]{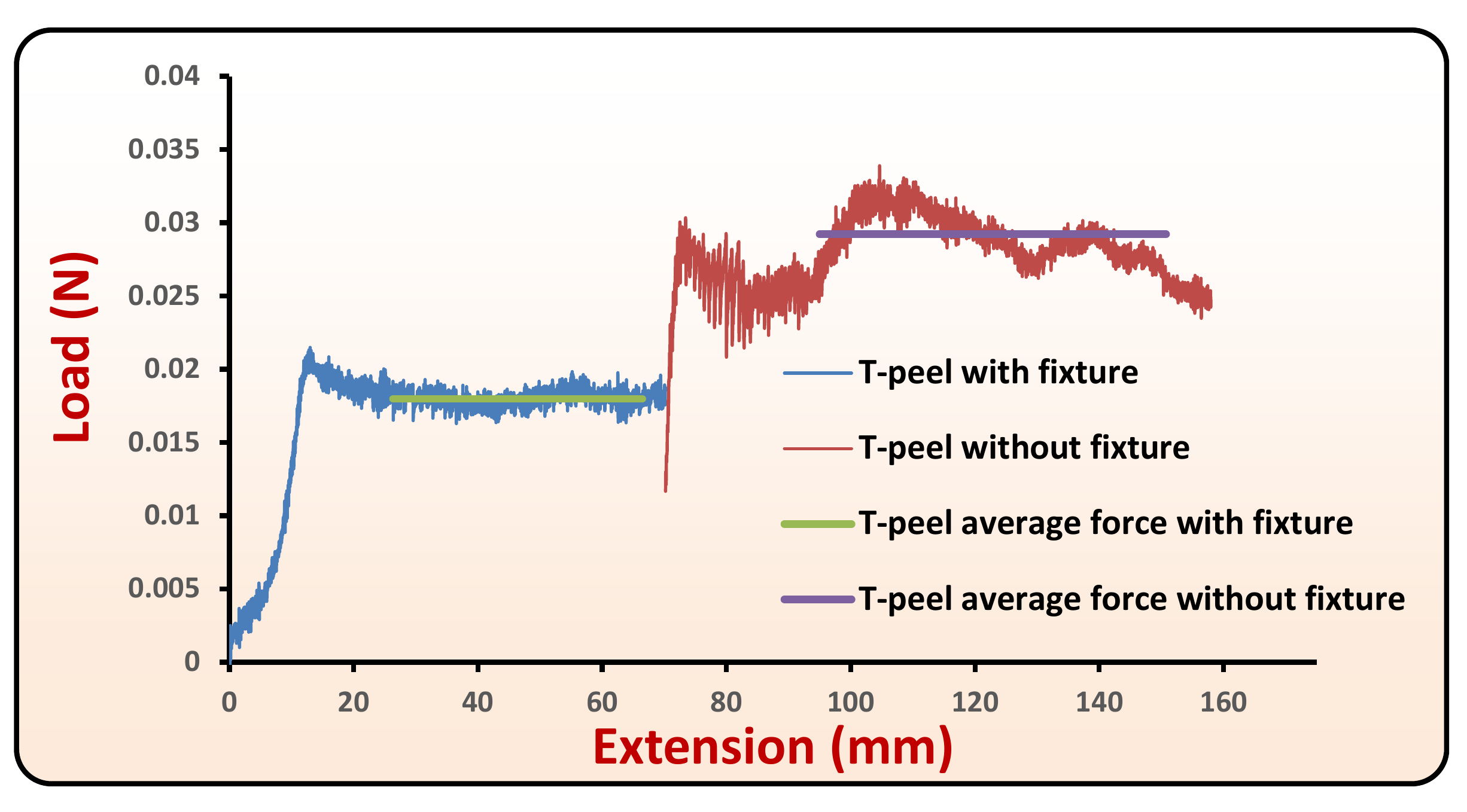}
\caption{T-peel test on a vertical machine with and without using the mechanical fixture for roll-bonded laminates sample `C'.}
\label{fig:S2-data-cropped}
\end{figure}

\textbf{Methods and Materials}\\

We considered several flexible adhered systems and, conducted T-peel tests on each system with and without using
the peel fixture. The following adhered systems were chosen: (1) three roll-bonded laminates, namely, sample `A' ($14.3$ mm wide), sample `B' ($13.34$ mm wide) 
and sample `C' ($13.1$ mm wide). The specimens were made from Hydroxypropyl Methylcellulose (HPMC) of grade E 15 and Polyethylene glycol (PEG)-400, with 
42.3\% (Wt. \%) of PEG in films. The samples `A',`B' and `C' were roll-bonded at a nominal plastic strain of 14.07\%, 23.3\% and 27.5\%, respectively. 
For complete details see \cite{padhyePhD2015,padhye-roll-bonding-machine,padhye-nature-bonding}, 
(2) two bonded layers of a commercial duct tape ($12.7$ mm wide), (3) two bonded layers of a Scotch brand cellophane tape commonly used in the office ($6.35$ mm wide), 
(4) two layers of Teflon glued together with EVA ($12$ mm wide), (5) two layers of Teflon glued together with a silicone encapsulant adhesive ($13.2$ mm wide), 
(6) two layers of self-adhering commercial vinyl laminates ($13.25$ mm wide), (7)
two layers of self-adhering commercial vinyl laminates (with backing 1 and $12.4$ mm wide), and (8) two layers of self-adhering commercial vinyl laminates (with backing 2
and $12.9$ mm wide). 
Peel tests were performed at a cross-head speed of 15 mm/min.

For all adhered systems, the experiment was first initiated in presence of the fixture to support the specimen's tail end, 
and corresponding force versus displacement curves were recorded until a steady-state was achieved. 
Once a sufficient amount of data was sampled in the steady state with the fixture, the experiment was paused,
the fixture was disengaged, the support for the tail end of the specimen was removed, and the test was re-initiated from 
the same point onwards while recording the corresponding load versus displacement curves. 
By removing the support, flexible adherends experienced a significant asymmetry due to the 
action of gravity. 

\begin{figure}[!htb]
\centering
\includegraphics[scale=0.5]{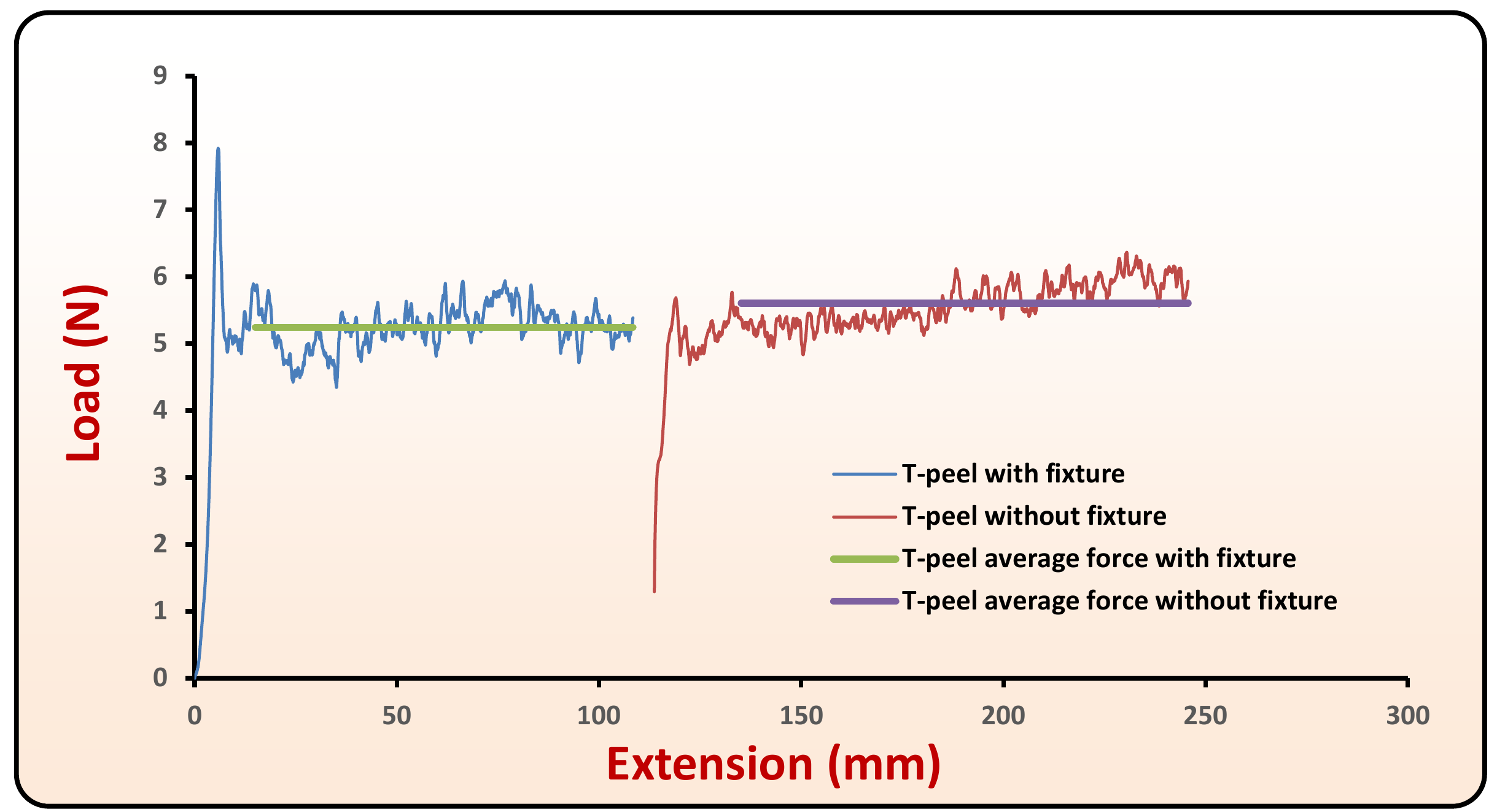}
\caption{T-peel test on a vertical machine without using the mechanical fixture for duct tape.}
\label{fig:t-peel-test-comparison-on-black-duct-tape}
\end{figure}

\begin{figure}[!htb]
\centering
\includegraphics[scale=0.5]{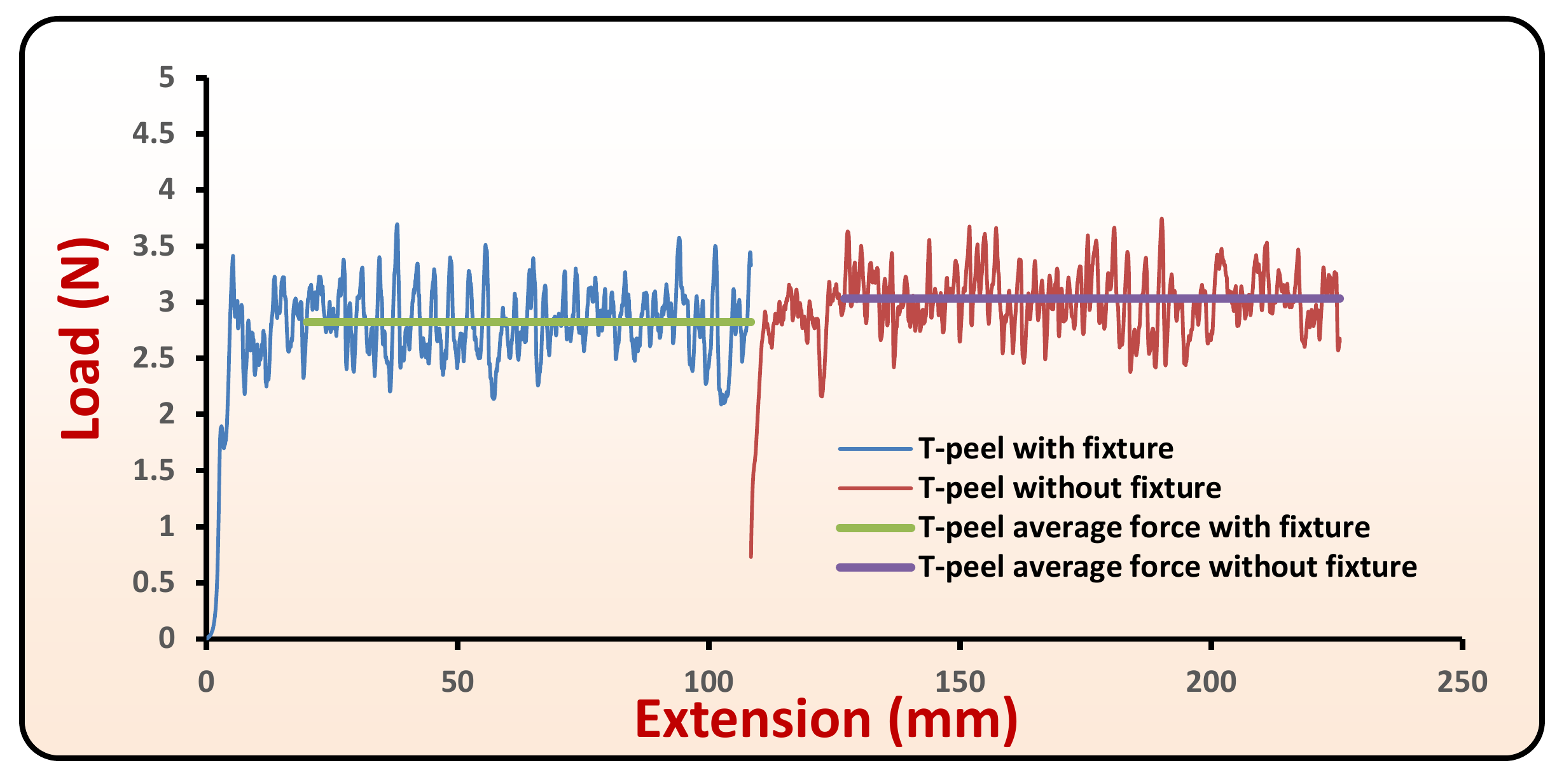}
\caption{\label{fig:t-peel-test-comparison-on-cellophane-tape} T-peel test on vertical machine with and without using the mechanical fixture for cellophane tape.}
\end{figure}

\begin{figure}[!htb]
\centering
\includegraphics[scale=0.5]{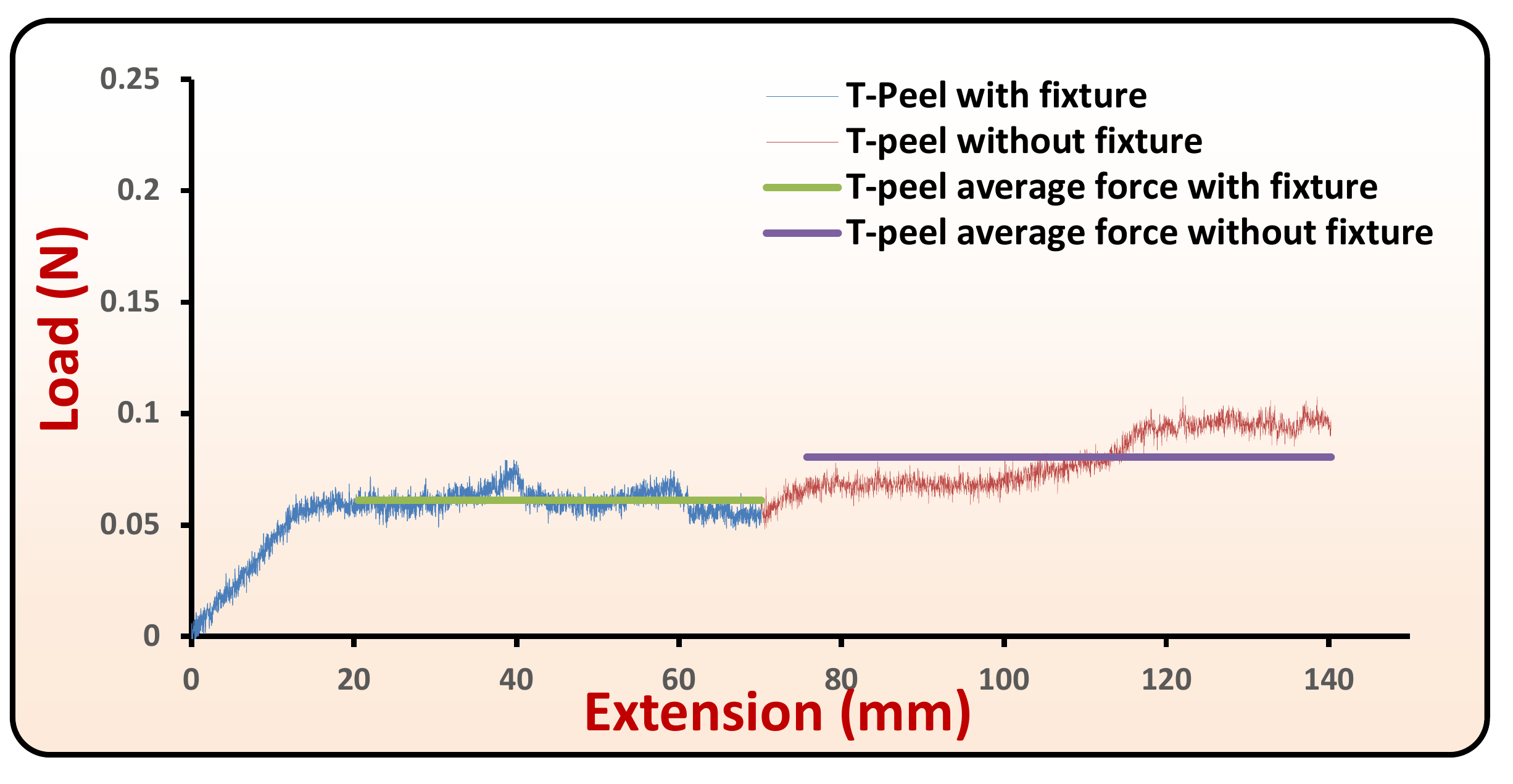}
\caption{T-peel test on a vertical machine with and without the fixture for Teflon adhered with EVA.}
\label{fig:2B2-cropped}
\end{figure}

\begin{figure}[!htb]
\centering
\includegraphics[scale=0.5]{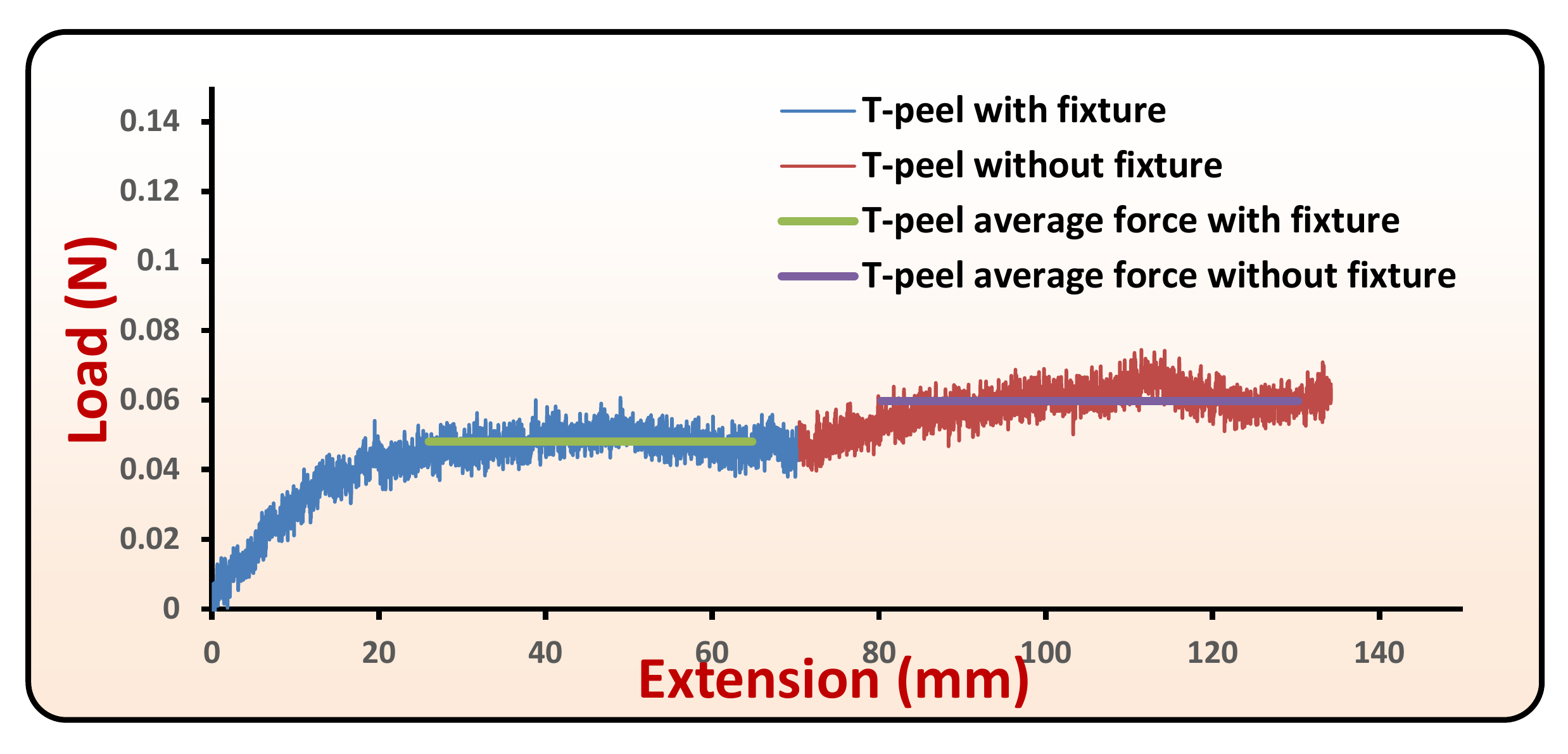}
\caption{T-peel test on vertical machine with and without using the fixture for Teflon adhered with a silicone encapsulant adhesive.}
\label{fig:2A2-cropped}
\end{figure}
\begin{figure}[!htb]
\centering
\includegraphics[scale=0.5]{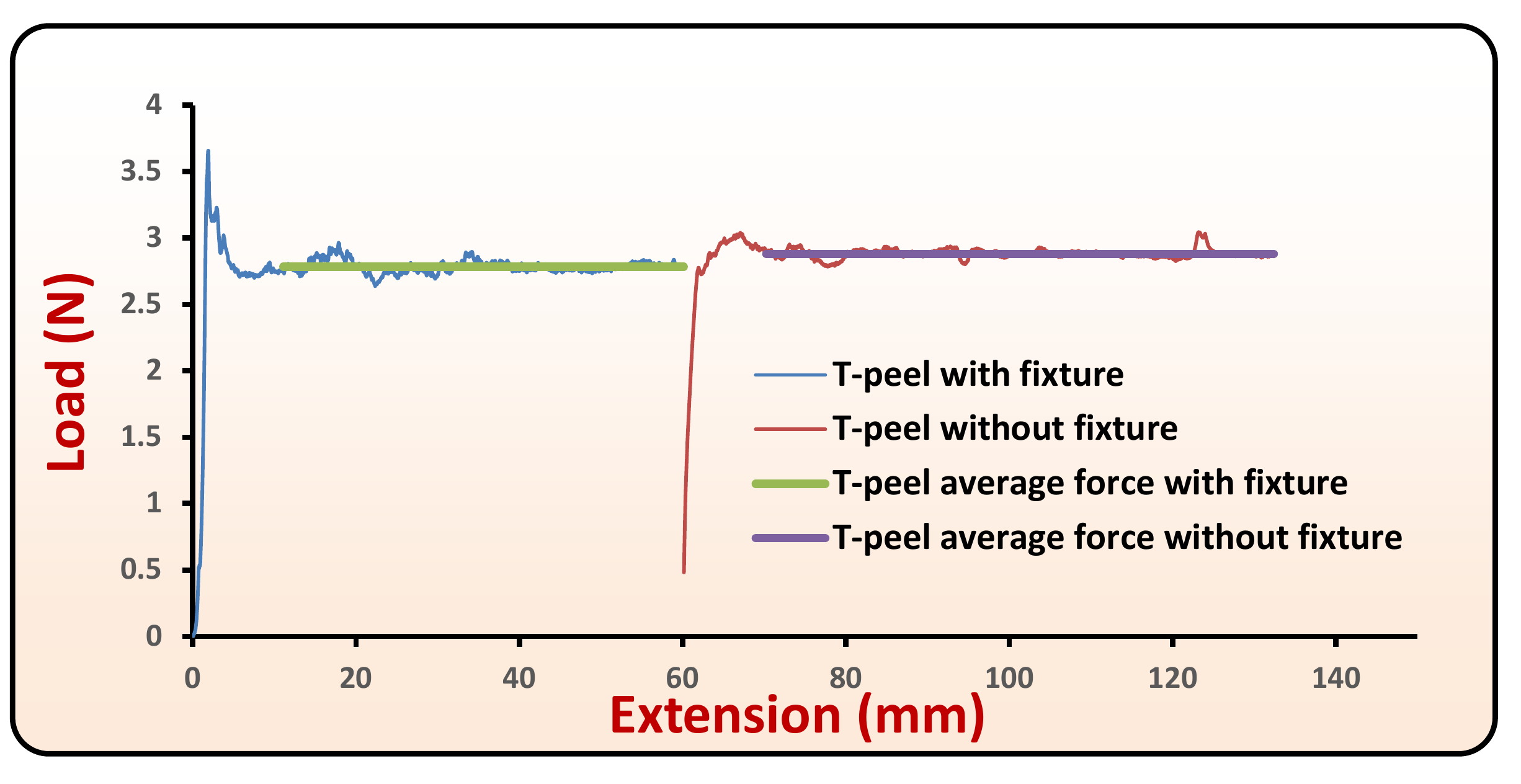}
\caption{T-peel test on vertical machine with and without using the fixture for self-adhering commercial vinyl laminates.}
\label{fig:3D3-cropped}
\end{figure}
\begin{figure}[!htb]
\centering
\includegraphics[scale=0.5]{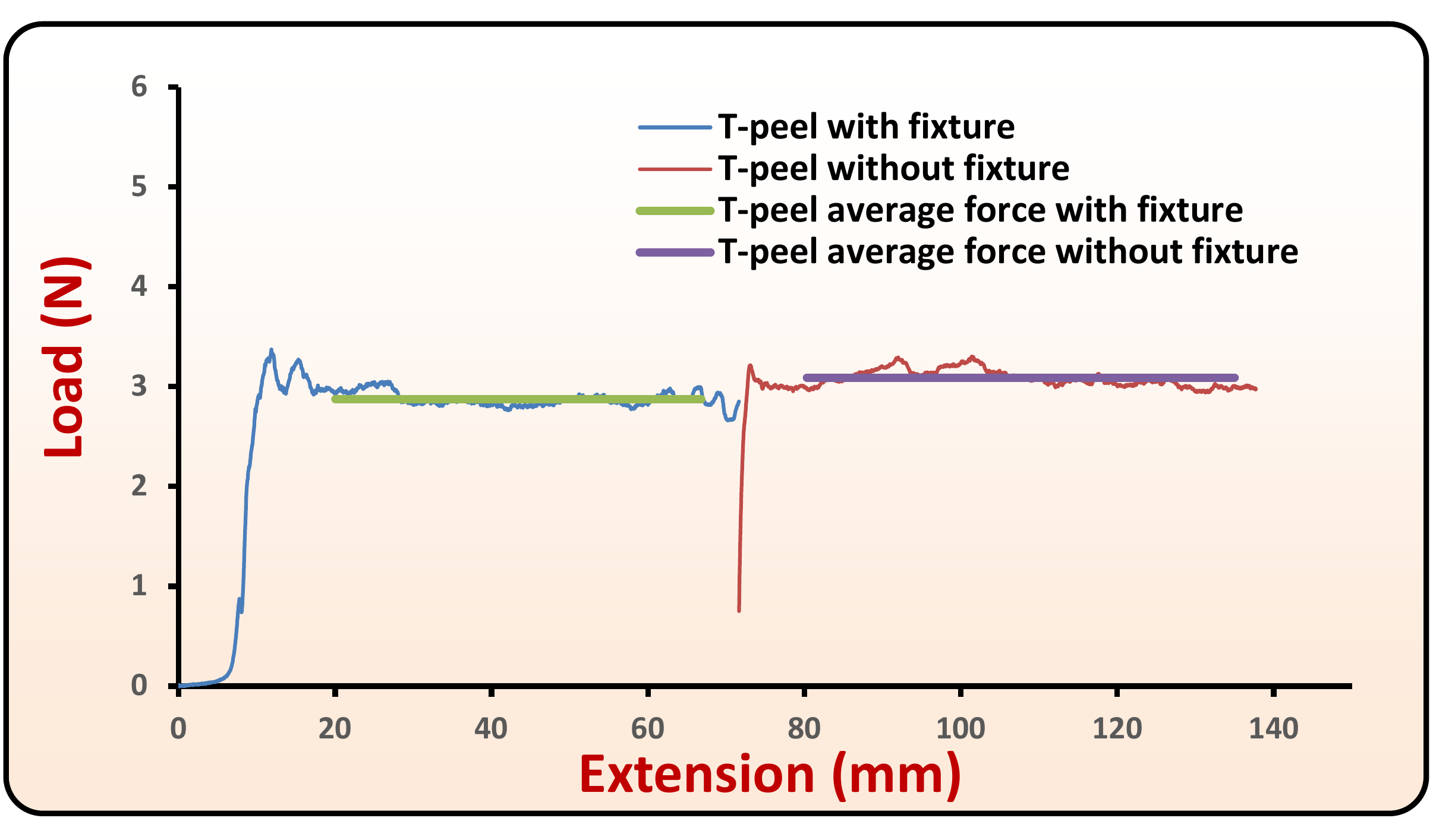}
\caption{T-peel test on vertical machine with and without using the fixture for self-adhering commercial vinyl laminates (with backing 1).}
\label{fig:3dD3d-cropped-II}
\end{figure}
\begin{figure}[!htb]
\centering
\includegraphics[scale=0.5]{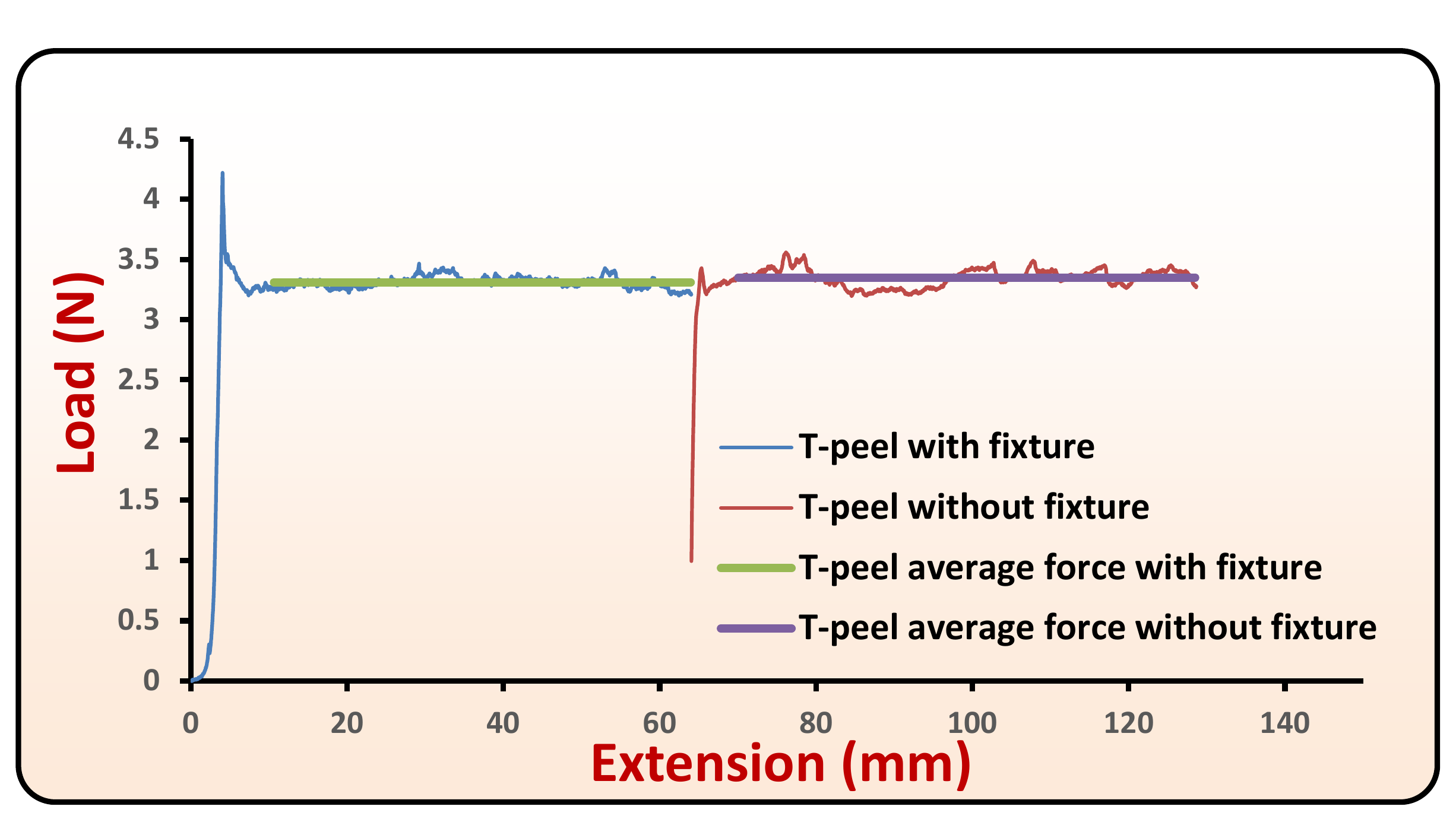}
\caption{T-peel test on vertical machine with and without using the fixture for self-adhering commercial vinyl laminates (with backing 2).}
\label{fig:3eD3e-cropped}
\end{figure}

\begin{table*}
 \begin{center}
 \caption{\textbf{Summary of the experimental results}}
 \label{tab:summary-of-experimental-results}
    \begin{tabular}{ |l|l|l|}
    \hline
\textbf{S. No.}&	\textbf{T-peel Specimen}						 	&	\textbf{\% Overestimation}$^*$\\ \hline \hline
1			&	Roll Bonded Laminates Sample `A' $\;\;$					&	85.65			\\
2			&	Roll Bonded Laminates Sample `B' $\;\;$					&	54.66			\\
3			&	Roll Bonded Laminates Sample `C' $\;\;$					&	62.48			\\
4			&	Duct-Tape 										&	6.88			\\
5			&	Cellophane-Tape									&	7.35		  			\\
6			&	Teflon-EVA-Teflon									&	31.79	   	      			\\
7			&	Teflon-silicone-encapsulant-Teflon					&	24.41  		       			\\	
8			&	Vinyl-Vinyl										&	3.51							\\  
9			&	VinylLaminate1-VinylLaminate1						&     7.46						\\  
10			&	VinylLaminate2-VinylLaminate2						&	1.19							\\	\hline
					\multicolumn{3}{l}{} 	\\
		\multicolumn{3}{l}{     $^*$ \textbf{\%Overestimate} = $\frac{\; Avg. \; Force\;  without\;  Fixture - Avg. \; Force \;  with\;  Fixture \; }{Avg.\;  Force\;  with \; Fixture}$} 
    \end{tabular}
 \end{center}
\end{table*}

Figure ~\ref{fig:peel-test-with-mechanism} shows a snapshot of the T-peel test on roll-bonded laminates sample `A' where the gravity effect 
has been suppressed due to the base support of the mechanical fixture. Figure ~\ref{fig:Without-mechanism-peel-test} shows 
the peel test on the same specimen without the support, such that action of gravity is dominant and introduces significant asymmetry.
The corresponding peel force versus displacement curves for the two cases are shown in Figure ~\ref{fig:t-peel-test-comparison}.
The average peel force was calculated in the steady state region by neglecting the initial or final peaks (or jumps)
and plotted with a solid line over the corresponding range. In the case of roll-bonded laminates sample `A', 
the average peel force in the asymmetric case is significantly (approximately 85.65\%) higher than the 
average peel force with the use of the fixture. The peel force versus displacement curves 
for roll-bonded samples `B' and `C' are shown in Figures ~\ref{fig:S1-30th-aug-2015-cropped} and ~\ref{fig:S2-data-cropped}, respectively,
and also show an increased unsteady peel force when tested without the fixture.
As mentioned above, the notion of steady state in the asymmetric case is ill-defined
because the overhanging length of the tail specimen decreases as the test proceeds; however, the average in the peel force is still 
calculated in the asymmetric regime.
We emphasize that a larger peel force (measured by the load cell at the
upper gripping end), when tested without the fixture can be attributed to
the mixed-mode failure, the specimen weight, possible plastic bending, etc., and 
there is no straight-forward method of precisely quantifying the contribution of
each factor in the increased peel force. 
If we the consider the case
of roll bonded specimen case `A', with a total length (L) of $200$ mm, 
width (w) of $14.3$ mm, thickness (2t) of $0.58$ mm and density ($\rho$) of 1180 Kg/m$^3$,
then the specimen weight due to the upper peel arm ($\rho$Lwt$g$) works out to be  $0.009$ N. The peel forces in the
symmetrical and asymmetrical cases were $0.028$ N and $0.052$ N, respectively. Thus,
the increase in peel force, approximately $0.024$ N, is greater than the weight of
the upper peel arm. Since the increase in the peel force, from symmetrical to assymetrical
case, is not alone due to the weight of the peel arm clamped at the load sensing grip, 
no straight-forward correction is possible. Also, in our experimental results, 
we achieved a steady state force in the symmetrical phase
with relatively short length of peel arms, and the weight of the 
part of peel arm which was unsupported by the fixture was noted to be almost an order of magnitude lower than the steady-state peel force.
This indicates that our proposed symmetrical test is quite close to the idealized T-peel test (symmetrical test in the absence of gravity). A detailed discussion on the effect of the weight of the peel arm in the symmetrical T-peel is given in Section ~\ref{sec:effect-of-weight}.

In the asymmetric peeling, a low modulus and thin film is likely to bend more, due to which the mode II component will increase.
As the mode II component increases, the effective toughness is likely to increase \cite{de2008modeling-bak,reeder1990mixed}. 
Note that the load versus displacement curves in the asymmetric case also show a slight lowering trend during the asymmetric 
part of the test, and the following reasons could possibly account for the lowering in peel force: (i) the total energy of an elastica is finite and constant, and in the symmetric case, 
the tail of the specimen carries no elastic energy; therefore, the rate of work performed by external force
exactly balances the interface toughness. However, in the asymmetric case, the tail end bends due to its own 
weight and possess some elastic energy. As the tail is drawn in, it releases its elastic energy, which would be
available for cracking. (ii) It is evident that greater asymmetry leads to more pronounced component 
of mode II; thus, when the tail end is long and asymmetry due to bending is high, the degree of 
interfacial shearing is dominant with respect to interfacial opening. However, as the tail shortens by being drawn into the crack tip, the
mode II component decreases, and therefore, the peel force decreases.  
In our experiments, it was also observed that the light weight samples, when tested without the fixture, 
suffered noticeable dynamic instability due to air currents, i.e., the samples were observed to shake and sometimes flutter.  
Such instabilities could also introduce undesired mode III parasitic effects of failure, and contribute to 
the increase in the peel force. This could also explain the rising trends in the recorded peel forces during the asymmetric testing. 
The use of support plate prevented such instabilities from occurring, and a very good steady state 
was noted during the symmetric part of the test. Further design modifications can be made to the support 
plate to eliminate any further unwanted effects of air-current on the testing of extremely delicate
and highly sensitive samples. An example would be the presence of a channel on the support plate 
where the specimen tail can reside. The summary of \% overestimation in the average peeling force with 
and without using support, are listed in Table ~\ref{tab:summary-of-experimental-results}.

In other adhered systems, as shown in Figures 
~\ref{fig:t-peel-test-comparison-on-black-duct-tape} to ~\ref{fig:3eD3e-cropped},
an increased average peel force is observed when the mechanism support is not used. 
For these specimens, the estimate of the overhanging length of the tail in the asymmetric-half of the test can be
made from the extension (cross-head displacement) during that phase of testing.

Commonly available duct and clear cellophane tapes, Figures ~\ref{fig:t-peel-test-comparison-on-black-duct-tape} and ~\ref{fig:t-peel-test-comparison-on-cellophane-tape}, exhibit noticeable fluctuations in the peeling force. 
Such fluctuations in the peeling force could be attributed to the micro-mechanisms of the fracture process itself, for example, instabilities at the crack tip,  
occasional slowing, and stopping of a crack shortly after the crack initiation.
Even if the geometry and loading are symmetrical, and 
macroscopically pure opening of the interfaces is noted, yet the crack path along the inter-adhesive
layer may not be in pure mode I microscopically. However, if one limits attention to a macroscopic viewpoint,
then failure can be regarded as mode I.

Furthermore, inhomogeneities such as air bubbles or local unevenness 
in the adhered system can lead to local fluctuations in the peel forces. Weakly 
adhered Teflon laminates, Figures ~\ref{fig:2B2-cropped} and ~\ref{fig:2A2-cropped}, 
exhibited a noticeable increase in average peel force when tested without the fixture. For self-adhesive 
vinyl laminates, Figures ~\ref{fig:3D3-cropped}, ~\ref{fig:3dD3d-cropped-II}, and ~\ref{fig:3eD3e-cropped},
the increase in peel force, while testing without the fixture, was only marginal.  

\section{CONCLUSION}
\label{sec:conclusion-and-future-work}

The peel test is a popular mechanical test to measure adhesive fracture energies for flexible laminates and 
adopted widely. In this paper, we have analyzed the T-peel test on vertical test machines and report
that effects of gravity
can lead to a significant rise in experimentally measured peel forces in comparison
the T-peel test where symmetry is maintained. 
To address the undesired effects arising due to the gravity, a mechanical 
fixture was developed, which can be readily adopted to different test machines. The fixture has a support plate that is kinematically
coupled, through a set of racks and gears, with the cross-head of 
the mechanical tester. The support plate moves at half the speed of the upper
cross-head, thus maintaining symmetry in the T-peel specimen and suppressing
the effects of specimen weight, mixed mode failure etc. 
T-peel test results on several adhered systems, with and without the use of mechanical peel-fixture,
demonstrate the importance of providing a symmetrical support to the specimen tail during peeling. This design thus greatly improves the performance of T-peel testing, 
and an accurate estimation of mode I fracture toughness (G$_{Ic}$). 
As an ongoing study, we are parametrically varying several non-dimensional quantities based on the 
length of overhanging tail, peak stresses in opening and shearing mode, ratio of interface toughness in mode I mode II,
yield strength, modulus, density, specimen geometry, etc. to explore which factors are dominant under what conditions and 
identify the underlying insightful trends of the asymmetric peeling.

\section{Acknowledgments}
This research was supported by Novartis Pharma AG and carried out at the NVS-MIT Center for Continuous Manufacturing Laboratories. 

\section{Appendix}
\label{sec:appendix}

Here, we present the mechanics of the peel test, both in symmetric and asymmetric case, based on slender beam theory. 
The slender beam theory assumes that stresses arise due to the bending moments,
and the role of axial or shear forces is negligible.

\subsection{Symmetric (ideal) T-peel}
\label{sec:slender-beam-theory} 

Thin (or slender) members usually demonstrate small bending stiffness, and therefore, they can undergo 
large deflections and rotations with only small strains. Such one-dimensional members are also referred  to as an ``elastica''.
The first analytical solution to an elastica employing Euler-Bernouli beam bending equations
was presented in \cite{bisshopp1945large}. 

\begin{figure}[htp]
    \centering
\includegraphics[scale=0.35]{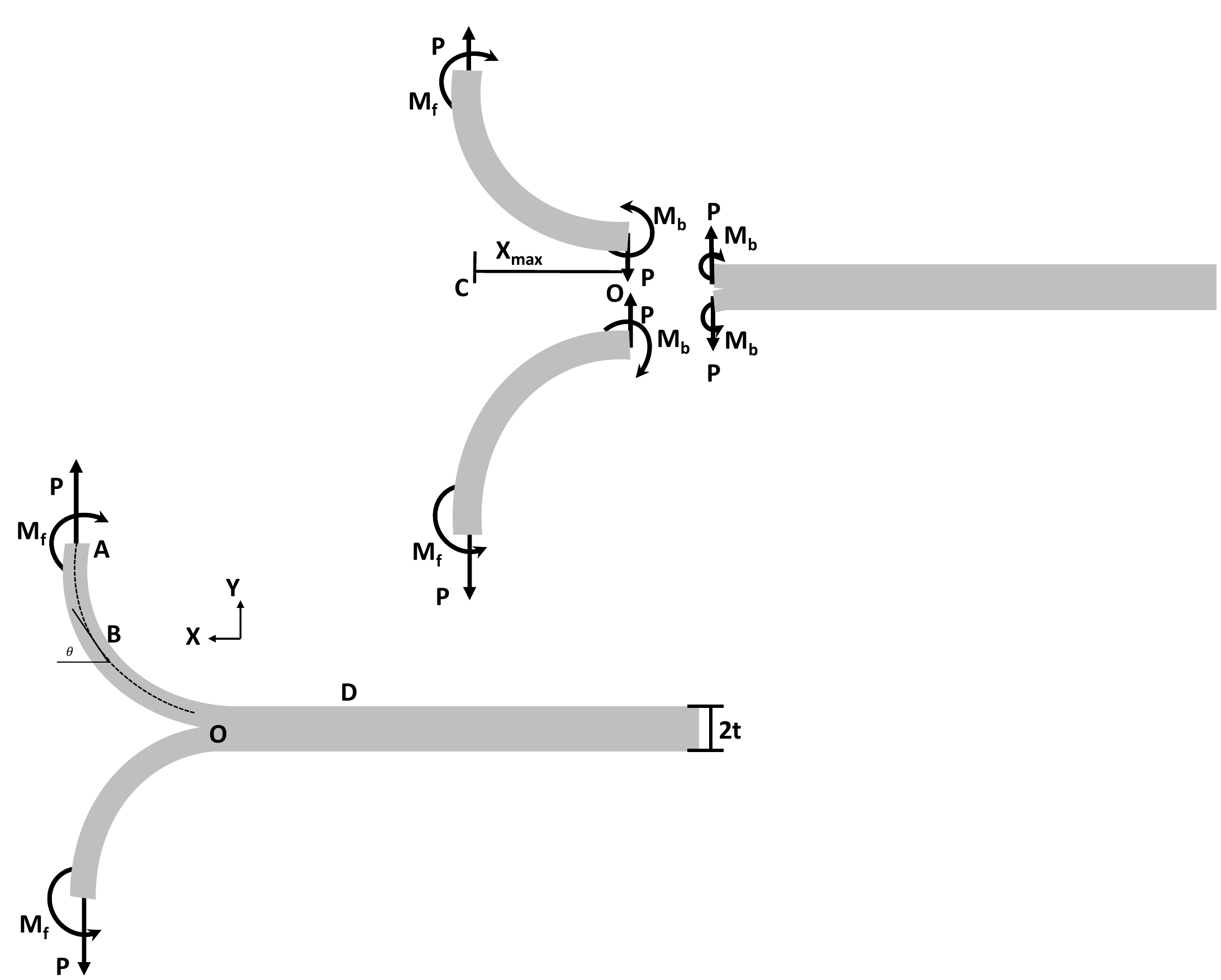}
    \textbf{\caption{ \label{fig:elastica-mechanics} Schematic of an ideal T-peel test. }}
\end{figure}

Figure ~\ref{fig:elastica-mechanics} shows the schematic of the ideal T-peel test. Here, the specimen width 
into the plane is considered to be large such that plane strain conditions hold.

In Figure ~\ref{fig:elastica-mechanics}, the crack tip location is marked at the location O. 
The upper peel arm is shown as OA, along with the forces and moments acting on it, and
due to symmetry the same holds for the lower peel arm. 

Adapting the elementary beam bending theory to the plane strain scenario, the relationships among 
strain, stress, radius of curvature, and moment are given as:
\begin{eqnarray}
\label{eqn:plane-strain-epsilon-y}
\epsilon_x=-\frac{y}{\rho}
\end{eqnarray}
\begin{eqnarray}
\label{eqn:plane-strain-sigma-x}
\sigma_x= E'\epsilon_x	=	-\frac{E'y}{\rho} = -E'y\frac{d\theta}{d s}		=	- \frac{My}{I}
\end{eqnarray}

Here we assume that the elastica is inextensible. 
The moment equilibrium of a beam element, say at point B, implies
\begin{eqnarray}
\label{eqn:1}
E'I\frac{d\theta}{d s}	= P(x_{\mathrm{max}}-x)+M_f
\end{eqnarray}
Differentiating the above equation with respect to arc length \textit{s}, we get
\begin{eqnarray}
\label{eqn:2}
E'I\frac{d^2\theta}{d s^2}	= - P\frac{dx}{ds}.
\end{eqnarray}
Using the fact that
\begin{eqnarray}
\label{eqn:3}
ds\;\cos(\theta)	= dx
\end{eqnarray}
and 
\begin{eqnarray}
\label{eqn:4}
ds\;\sin(\theta)	= dy,
\end{eqnarray}
and substituting equation ~\ref{eqn:3} in equation ~\ref{eqn:2}, we get
\begin{eqnarray}
\label{eqn:5}
E'I\frac{d^2\theta}{d s^2}	= - P \cos(\theta).
\end{eqnarray}
Re-arranging, we get
\begin{eqnarray}
\label{eqn:6}
\frac{d^2\theta}{d s^2} + k \cos(\theta)	=	0,
\end{eqnarray}
where $k=\frac{P}{E'I}$. Now, using a substitution $\frac{d \theta}{d s}=v$ and integrating the above equation, 
we get\\
\begin{eqnarray}
\label{eqn:7}
\frac{v^2}{2} + k \sin(\theta)	=	c_1,
\end{eqnarray}
where $c_1$ is the constant of integration, which can be obtained through boundary conditions 
on $v$, i.e., $\frac{d \theta}{d s}$. Now, using equations ~\ref{eqn:3}, ~\ref{eqn:4}, and ~\ref{eqn:7}, we can find

\begin{eqnarray}
\label{eqn:8}
S=\int_0^s ds = \int_{0}^{\theta}	\frac{d \theta}{\sqrt{2(c_1-k \sin(\theta))}};
\end{eqnarray}
\begin{eqnarray}
\label{eqn:9}
X=\int_0^x dx = \int_{0}^{\theta}	\frac{\cos(\theta) d \theta}{\sqrt{2(c_1-k \sin(\theta))}};
\end{eqnarray}
\begin{eqnarray}
\label{eqn:10}
Y=\int_0^y dy = \int_{0}^{\theta}	\frac{\sin(\theta) d \theta}{\sqrt{2(c_1-k \sin(\theta))}}.
\end{eqnarray}

In this treatment, we have considered the unattached part as a beam whose tangent rotates from $\theta=0$ to 
$\theta=\pi/2$, and thus, we have ignored base rotation effects. In general, $\theta=0$ may not be true, i.e., $\theta=0$ may not
mark the onset of the unbonded part of the peel arm. The value of $\theta$ at the base depends on the 
cohesive zone extension between the two adherends \cite{cotterell2006root,williams1993root}.
Although, the extent of the cohesive zone at the crack tip is problem-dependent, as long as the symmetry is maintained, the region ahead of the crack tip 
comprises only of normal stresses (and no shear stresses). 
Equations ~\ref{eqn:9} and ~\ref{eqn:10} also have an analytic solution when $\theta=0$ at the base \cite{kim1988elastoplastic}. 

Y as well as S can grow unbounded depending on the boundary conditions. If we assume the peel arm to be long while staying within
the elastic limits, then boundary conditions as per Figure ~\ref{fig:elastica-mechanics} can be set as
$M_f=0$ at $\theta=\pi/2$ or the curvature $\kappa_b=\frac{d\theta}{ds}=0$ at $\theta=\pi/2$.  Thus, 
using equation ~\ref{eqn:7}, we get
$$
c_1=k=\frac{P}{E'I}
$$ 
Integrals in equations ~\ref{eqn:8}, ~\ref{eqn:9}, and ~\ref{eqn:10}, reduce to\\
\begin{eqnarray}
\label{eqn:11}
S_{\mathrm{total}}=\int_0^s ds = \int_{0}^{\pi /2}	\frac{d \theta}{\sqrt{2k(1-\sin(\theta))}}
\end{eqnarray}
\begin{eqnarray}
\label{eqn:12}
X_{\mathrm{max}}=\int_0^x dx = \int_{0}^{\pi /2}	\frac{\cos(\theta) d \theta}{\sqrt{2k(1-\sin(\theta))}}
\end{eqnarray}
\begin{eqnarray}
\label{eqn:13}
Y_{\mathrm{max}}=\int_0^y dy = \int_{0}^{\pi /2}	\frac{\sin(\theta) d \theta}{\sqrt{2k(1-\sin(\theta))}}
\end{eqnarray}

Equations ~\ref{eqn:11} and ~\ref{eqn:13} 
represent improper integrals of second type, since the function to be integrated becomes unbounded in the specified limits. 
This implies that elastica becomes vertical only in an asymptotic sense under the absence
of any moment at the upper grip. However, for practical purposes, this is attained approximately quite easily. Numerical integration can also be employed
in the range [0, $\pi/2$) to solve for the above quantities. 

Next, we calculate the elastic strain energy stored in a part of elastica from an angle of $0$ to $\theta$.
For any differential element

\begin{eqnarray}
\label{eqn:14}
M=\frac{E'I}{\rho}=\kappa E'I;
\end{eqnarray}
\begin{eqnarray}
\label{eqn:15}
\alpha=\kappa ds.
\end{eqnarray}

Thus, energy stored in an element of length $ds$ in bending from an angle $0$ to $\alpha$ (or attaining a curvature $\kappa$) is given as follows:
\begin{eqnarray}
\label{eqn:16}
U_{\mathrm{element}}(\alpha)=\int dU_{\mathrm{element}}=\int_0^\alpha M(\alpha) d \alpha.
\end{eqnarray}
Using equations ~\ref{eqn:14} and ~\ref{eqn:15}, we can also write	\\
\begin{eqnarray}
\label{eqn:17}
dU_{\mathrm{element}}(\kappa)=E'I  \kappa \; d\kappa \; ds.
\end{eqnarray}
Thus, 
\begin{eqnarray}
\label{eqn:18}
U_{\mathrm{element}}(\kappa)= \int_0^\kappa E'I  \kappa \; d\kappa \; ds	= E'I \frac{\kappa^2}{2} ds.
\end{eqnarray}
Consequently, the energy of the part of elastica from O up to any point on it is given as follows:
\begin{eqnarray}
\label{eqn:19}
U_{\mathrm{net}}(s)= \int_0^s E'I \frac{\kappa^2(s)}{2} ds.
\end{eqnarray}
This can be re-written in terms of $\theta$ as follows:
\begin{eqnarray}
\label{eqn:20}
U_{\mathrm{net}}(\theta)= \frac{E'I}{2}\int_0^{\theta_f} \sqrt{2k(1-\sin (\theta))} \;\; d \theta
\end{eqnarray}

For $\theta_f=\pi/2$, U$_{\mathrm{net}}$ represents the total energy of the elastica. The total energy is always 
finite even though the length of the peel arm is unbounded. We repeat that this is true when elastica
is considered to be inextensible. 

Under steady-state conditions if the energy of elastica is constant (and no dissipative work occurs in the specimen), 
then the rate of external work equals the rate of work for interfacial de-cohesion. 

\begin{equation}
\label{eqn:external-work}
W_{\mathrm{external}}	= 2P\times \dot{x}=	G_c \times w \times \dot{x} 
\end{equation}

or

\begin{equation}
\label{eqn:GIC}
G_{Ic} 	=	\frac{2P}{w}
\end{equation}

An important aspect for experimental consideration is the length (L$_{\mathrm{unbonded}}$) of initial unbonded part for gripping
the specimen in the mechanical tester.  As per Figure ~\ref{fig:elastica-mechanics}, if in the initial 
phase of gripping the specimen in the mechanical tester only moments act the at the clamping end A, then M$_f$ = M$_b$ = $M'$ (say), and the unbonded 
part of the peel arm from A to O will form a circular quarter-arc with a constant curvature of $\kappa = \frac{M'}{EI}$. 
To avoid plastic deformation during clamping, the initial length should be chosen 
such that the initial radius of curvature $\rho$ is large enough to avoid plastic bending.
The radius of curvature is 

\begin{equation}
\label{eq:rho-length-relation}
\rho = \frac{2L_{\mathrm{unbonded}}}{\pi};
\end{equation}

Plasticity is avoided by choosing $L_{unbonded}$ such that $\rho$ is in the elastic limit, i.e.,

\begin{equation}
\label{eq:rho-elastic}
\rho > \frac{Et}{2\sigma_{\mathrm{yield}}}.
\end{equation}

By combining equations ~\ref{eq:rho-length-relation} and ~\ref{eq:rho-elastic}, we obtain

\begin{equation}
\label{eq:initial-debonded-for-elastic}
L_{\mathrm{unbonded}} > \frac{\pi E t}{4\sigma_{\mathrm{yield}}}
\end{equation}

The criterion of equation ~\ref{eq:initial-debonded-for-elastic} was employed while performing the peel tests. 
On the derivation of conditions for the onset of plasticity in the peel
arm during bending see \cite{kim1988elastoplastic}.

\subsubsection{Role of Specimen Weight in ``Symmetrical'' T-peel}
\label{sec:effect-of-weight}
\begin{figure}[!htb]
\centering
\includegraphics[scale=.75]{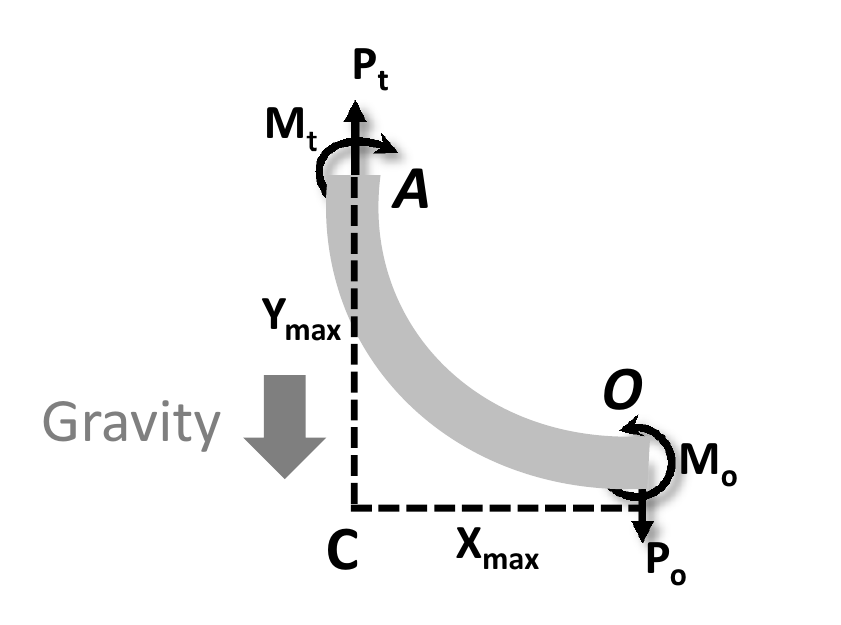}
\caption{Free body diagram of the upper peel arm under action of gravity during the symmetrical peel test.}
\label{fig:Elastica-role-of-weight-cropped-cantilever-beam}
\end{figure}

We wish to emphasize that in practice, although the use of the support plate leads to a symmetrical configuration 
visually and, largely suppresses the weight of the tail, the effect of gravity on the unsupported peel arms 
must be given a careful attention. Let us consider
the free body diagram of the upper peel arm `OA',
shown in Figure ~\ref{fig:Elastica-role-of-weight-cropped-cantilever-beam}, 
during symmetrical T-peel test with the gravity ``turned on''.

Here, $P_t$ is the peel force per unit width, $M_t$ is the 
moment per unit width at the upper clamping end, 
$M_o$ is the per unit width moment at the base end (which 
drives the opening of the crack), $P_o$ is the reaction force 
per unit width at 
the lower end, and $X_{max}$ and $Y_{max}$ denote the maximum dimensions
in $x$ and $y$ directions, respectively, with respect to the base $O$. 
Let $L$, $\rho$, $t$, $E'$ and $I$ denote the
length, density, thickness, plane strain modulus and moment of inertia of
the peel arm $OA$, respectively. 
For different levels of $P_t$, with $\theta$ equal to $0$ at the base $O$
and $\pi/2$ at the top end $A$, we wish to evaluate the quantities 
$M_o$, $M_t$, $P_o$, $X_{max}$ and $Y_{max}$ with gravity ``turned on'', and compare them to the ideal case with no gravity. 
In particular, we consider the plots of the following non-dimensionalized quantities, 
$\frac{M_o}{(E'I/Lb)}$, $\frac{M_t}{(E'I/Lb)}$, $\frac{P_o}{(E'I/L^2b)}$, $\frac{X_{max}}{L}$ and $\frac{Y_{max}}{L}$ with respect to the 
applied non-dimensionalized peel force $\frac{P_{t}}{(E'I/L^2b)}$. 
Commercial 
finite element program, Abaqus Standard, was employed for this purpose. 

\begin{figure}[!htb]
\centering
\includegraphics[scale=1.0]{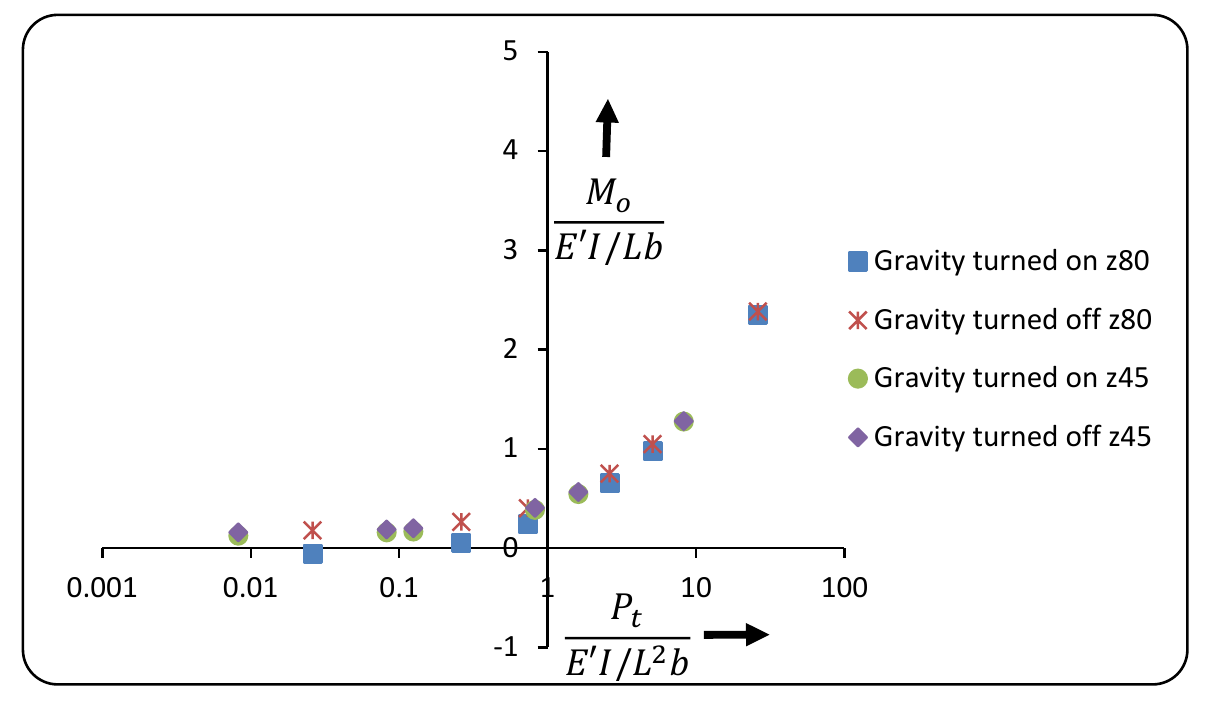}
\caption{Plot of non-dimensionalized moment at the base and 
peel force, with and
without gravity, for two different lengths (z45 and z80).}
\label{fig:New-ND-Plot-Mo_cropped}
\end{figure}

\begin{figure}[!htb]
\centering
\includegraphics[scale=1.0]{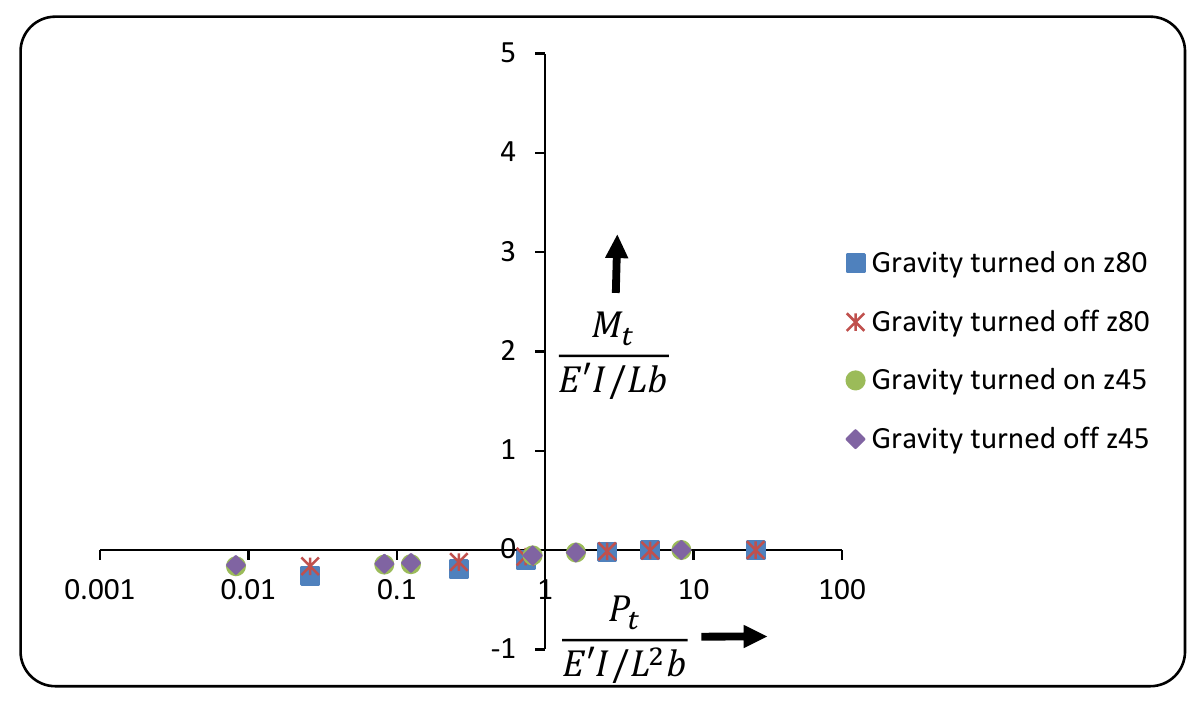}
\caption{Plot of non-dimensionalized moment at the top versus 
non-dimensionalized peel force, with and without gravity, for two different lengths (z45 and z80).}
\label{fig:New-ND-Plot-Mt_cropped}
\end{figure}

\begin{figure}[!htb]
\centering
\includegraphics[scale=.70]{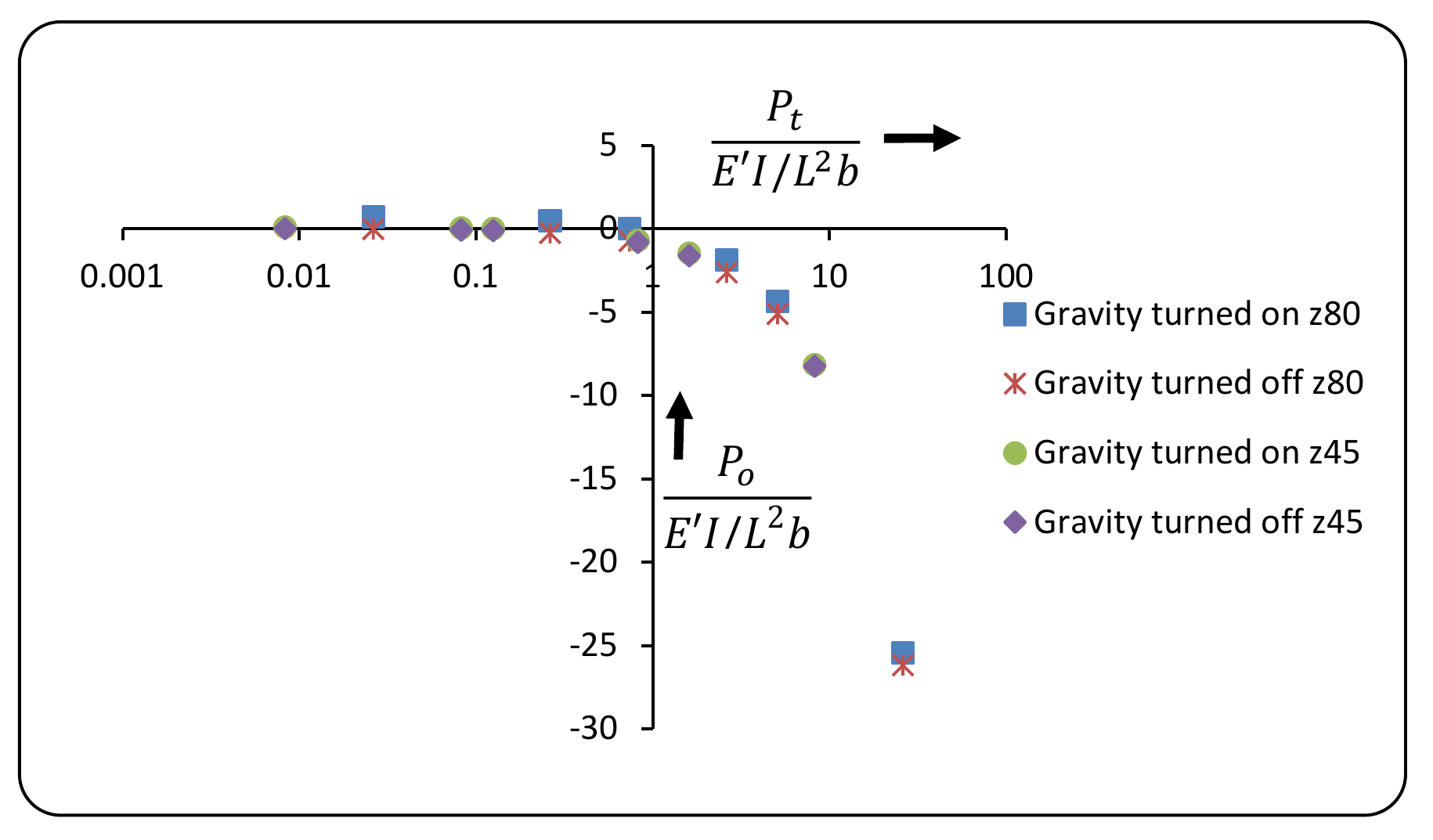}
\caption{Plot of non-dimensionalized reaction force at the base and non-dimensionalized peel force, with and without gravity, for two different lengths (z45 and z80).}
\label{fig:New-ND-Plot-Po_cropped}
\end{figure}

\begin{figure}[!htb]
\centering
\includegraphics[scale=.7]{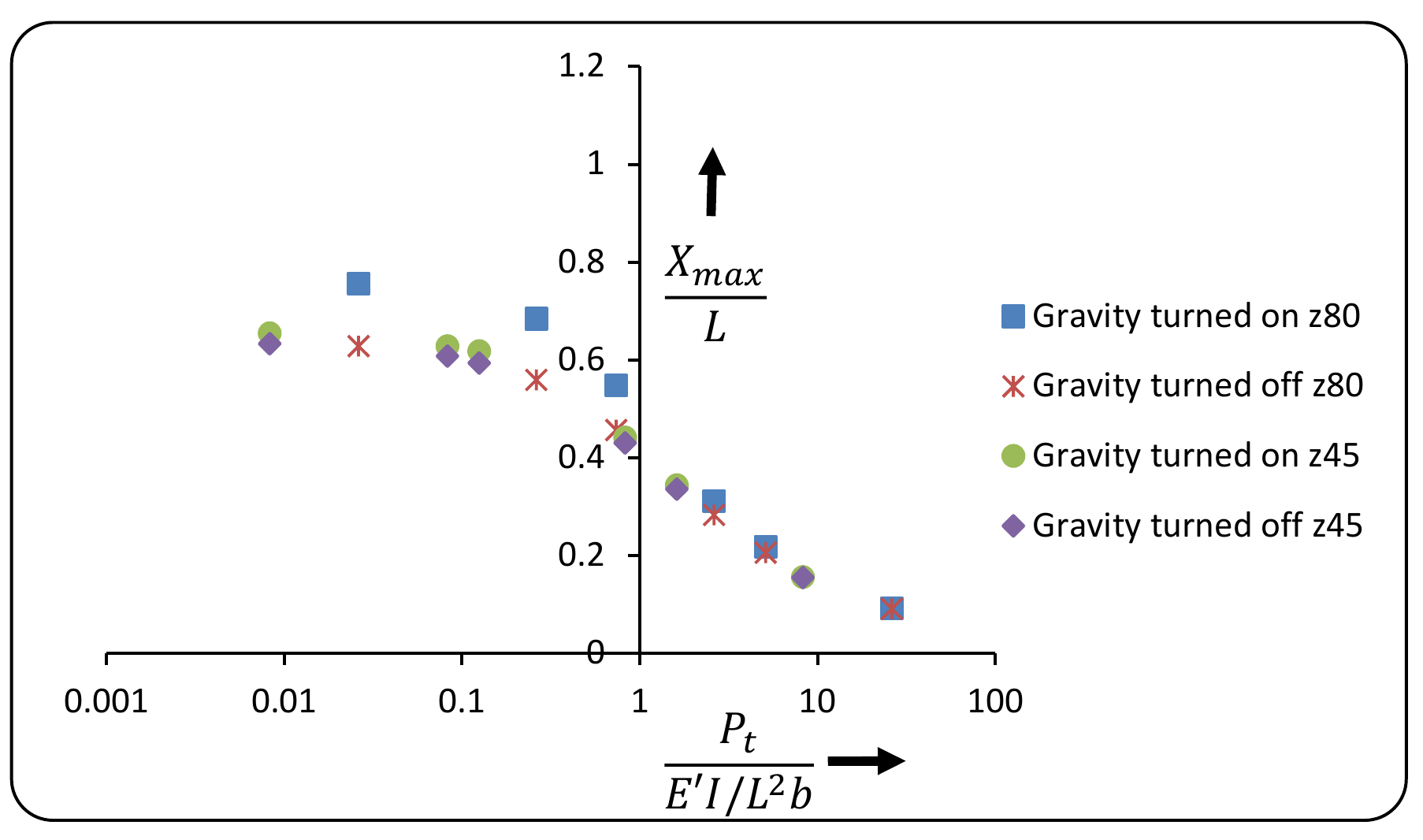}
\caption{Plot of non-dimensionalized maximum span in the x-direction versus
non-dimensionalized peel force, with and without gravity, for two different lengths (z45 and z80).}
\label{fig:New-ND-Plot-x_max_cropped}
\end{figure}

\begin{figure}[!htb]
\centering
\includegraphics[scale=.7]{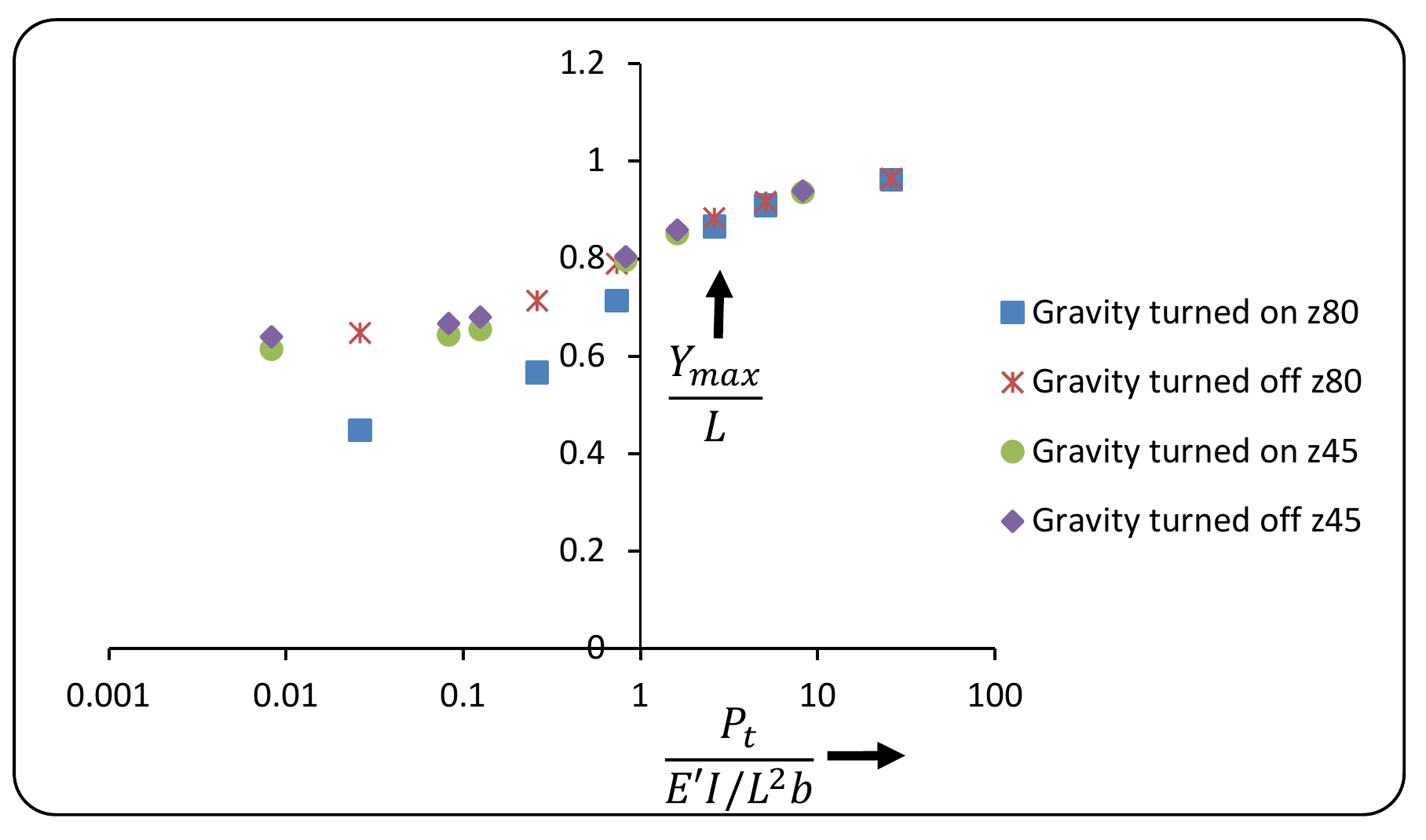}
\caption{Plot of non-dimensionalized maximum span in the y-direction versus
non-dimensionalized peel force, with and without gravity, for two different lengths (z45 and z80).}
\label{fig:New-ND-Plot-y_max_cropped}
\end{figure}

Typically, the initial clamped lengths of the peel arms employed in our experiments
were approximately $25$--$40$ mm. During the peel test, the weight of the 
supported part (unpeeled portion of the specimen) is continuously transfered to (and becomes the part of) 
the unsupported peel arms, and therefore it is possible to sense an increase in the peel force. 
Thus, for illustrations we choose two representative lengths of the peel arm 
$OA$ as $45$ mm and $80$ mm. A non-dimensional parameter 
$z=\rho gbtL^3/EI$ is evaluated at the two lengths 
$L=45$ mm and $L=80$ mm, and denoted as z45 and z80, respectively. 
$\rho$ and $t$ are chosen as 1180 kg/m$^3$ and 0.29 mm, respectively,
corresponding to the case (1), as discussed in section ~\ref{sec:results-and-discussions}. 
The plots of various non-dimensionalized quantities, with and without gravity,
for lengths of $45$ mm (z45) and $80$ mm (z80) are shown 
in Figures ~\ref{fig:New-ND-Plot-Mo_cropped}-~\ref{fig:New-ND-Plot-y_max_cropped}
 \footnote{Moments are considered
positive in anti-clockwise direction. Forces are considered negative in the downward direction}. 
Summarily, it is clear that difference between the quantities, with and without the gravity, decreases 
as the applied peel force increases (except for the reaction force $P_b$, where the difference
is constant and equal to the weight of the peel arm). 

In the experimental scenario corresponding to case (1), the levels of 
non-dimensionalized steady state peel forces were approximately 7.0-10.0 in magnitude,
and from the plots presented here; it is evident that the role of the weight of the peel arm is 
negligible at these levels of peel forces. For other adhered systems considered in this study, non-dimensionalized steady state peel forces
were noted to be much larger than 10.0 in magnitude and 
therefore the role of the gravity in 
all other cases was also negligible.

We conclude that when peel force per unit width is 
relatively high (up to an order of 
magnitude) compared to the normalizing factor $E'I/L^2b$ for these materials, 
then the effect of gravity on the peel arms can be simply ignored
in symmetric T-peel. 

However, we wish to emphasize that although the effect of gravity on lightweight peel
arms can be ignored during the symmetric peeling, it is possible that the \textit{total} 
sample weight can be comparable to the peel forces and may contribute significantly 
to an increased peel force during the asymmetric peeling (when the support plate is removed). 
It is expected that the illustrations presented in this article should facilitate 
the proper accounting of different factors contributing to the peel force (and the derived
fracture toughness).

\subsection{Asymmetric T-peel}

\begin{figure}[!htb]
\centering
\includegraphics[scale=.5]{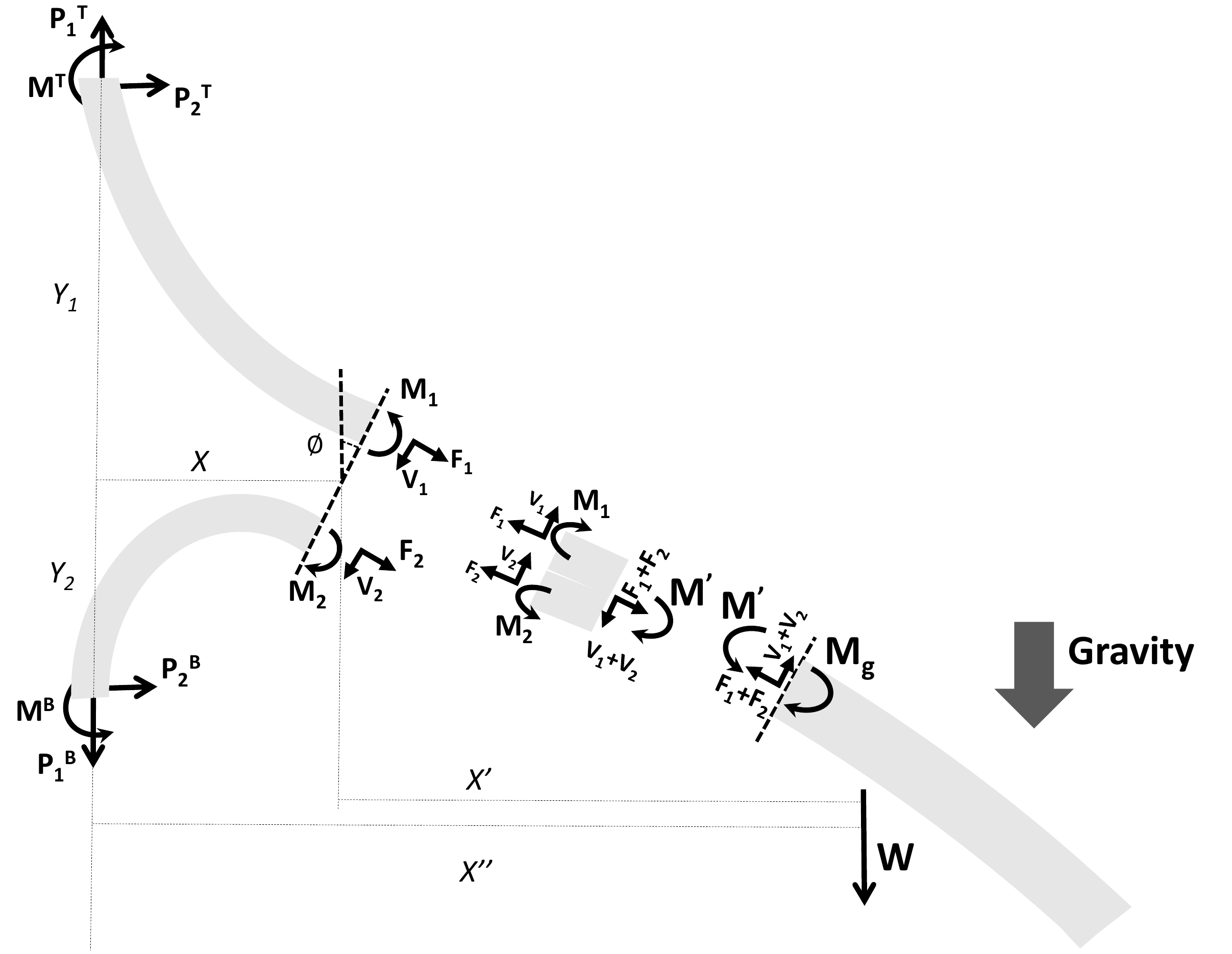}
\caption{Free body diagram of a beam under gravity.}
\label{fig:peel-test-gravity-hangover-freebodydiagram-cropped}
\end{figure}

Figure ~\ref{fig:peel-test-gravity-hangover-freebodydiagram-cropped} shows free body diagrams of the asymmetric peeling under the action of gravity. 

From the global equilibrium equation, we have

\begin{equation}
\label{eq:gravity-hangover-1}
P_1^T= P_1^B+W
\end{equation}

\begin{equation}
\label{eq:gravity-hangover-1b}
P_2^T+P_2^B=0
\end{equation}

If $M^{'}_g$ is the moment,
due to specimen weight,
about the axis through the 
line of action of vertical forces acting at the clamp, then the total moment equilibrium can be written as follows:

\begin{equation}
\label{eq:gravity-hangover-2}
M_T +	M^{'}_{g}  + P_2^T \times y_1=  M_B + P_2^B \times y_2
\end{equation}

Using equation ~\ref{eq:gravity-hangover-1b} in equation ~\ref{eq:gravity-hangover-2}, we have

\begin{equation}
\label{eq:gravity-hangover-3}
P_2^B=-P_2^T= \frac{(M_T-M_B+M^{'}_g)}{(y_1+y_2)}
\end{equation}

If we consider the free body diagram of the upper peel arm (while ignoring its own weight), and equilibrium conditions then we have

\begin{equation}
\label{eq:gravity-hangover-4}
P_1^T =  V_1 \times \cos \phi + F_1 \times \sin \phi;
\end{equation}

\begin{equation}
\label{eq:gravity-hangover-5}
P_2^T = V_1	\times \sin \phi - F_1 \times \cos\phi;
\end{equation}

\begin{equation}
\label{eq:gravity-hangover-6}
P_2^T\times y_1 + P_1^T\times x + M^T= M_1.
\end{equation}

Equilibrium equations for the specimen tail, of weight $W$, imply

\begin{equation}
\label{eq:gravity-hangover-7}
W\times \sin \phi =	F_1	+	F_2;
\end{equation}

\begin{equation}
\label{eq:gravity-hangover-8}
W\times \cos \phi =	V_1	+	V_2;
\end{equation}

\begin{equation}
\label{eq:gravity-hangover-9}
M_g= M'.
\end{equation}

Considering the equilibrium of lower peel arm (while ignoring its own weight), we have

\begin{equation}
\label{eq:gravity-hangover-10}
P_1^B	+V_2 \times \cos \phi + F_2 \times \sin \phi = 0;
\end{equation}

\begin{equation}
\label{eq:gravity-hangover-11}
P_2^B	=	V_2 \times \sin \phi - F_2 \cos \phi;
\end{equation}

\begin{equation}
\label{eq:gravity-hangover-12}
M_2			=	M_B			+	 P_2^B \times Y_2 +  P_1^B \times X;
\end{equation}

Considering the moment equilibrium of the intermediate section at the crack tip, we have

\begin{equation}
\label{eq:gravity-hangover-12b}
M_2			=	M_1			+	 M' + \triangle M
\end{equation}
where, $\triangle M$ is the resultant moment due to shear and normal forces acting on the intermediate
element. This implies

\begin{equation}
\label{eq:gravity-hangover-12b}
M_2		=M_1	+M_g + \triangle M.
\end{equation}

It is not necessary that the curvatures of three slender members are equal at their meeting point.  
Since we have only equilibrium equations for the system (two for the force balance and one for the moment balance), 
the system is statically indeterminate. We would need to consider compatibility, constitutive behavior, 
and kinematics to deterministically solve for all variables. 

Based on the free body diagram shown in Figure ~\ref{fig:peel-test-gravity-hangover-freebodydiagram-cropped},
the relation between the moments and forces acting on the intermediate section and interface toughness 
(G$_{Ic}$ and G$_{IIc}$ at the crack tip) is not straightforward. From the forces and 
moments acting on the intermediate element, analytical determination of crack tip stress intensities is not possible. 
In addition, during peeling, the equilibrium configuration for the entire system changes continuously, and 
therefore, deriving a correction for the associated effects is a major task requiring extensive numerical modeling. In fact, there will never be a 
steady state in the asymmetric case, since the degree of mode-mixity
will vary with the changing length of the overhang specimen tail. 


\input{rsitemplateAHSaug13.bbl}
\end{document}

%% file: rsitemplateAHSaug13.bbl
\providecommand{\noopsort}[1]{}\providecommand{\singleletter}[1]{#1}%